\documentclass[useAMS,usenatbib]{mn2e}

\def \isc{\textsc{i}}
\def \iscs{\textsc{i} }

\usepackage{ifpdf}
\usepackage{graphicx}
\usepackage{amsmath}
\usepackage{amsfonts}
\usepackage{amssymb}
\usepackage{times}
\usepackage{array}
\usepackage{supertabular}
\usepackage{url} 
\usepackage[figuresright]{rotating}
\usepackage{array}
\usepackage{caption} 
\usepackage{afterpage}
\usepackage{fixltx2e} 
\usepackage{multirow}

\newcolumntype{:}{>{\global\let\currentrowstyle\relax}}
\newcolumntype{;}{>{\currentrowstyle}}

\title[New constraint on $\mu$-variation from Q0528$-$250]{New constraint on cosmological variation of the proton-to-electron mass ratio from Q0528$-$250}
\author[J. A. King et al.]{Julian A. King$^{1,3}$\thanks{E-mail:
jking.phys@gmail.com}, Michael T. Murphy$^{2}$\thanks{E-mail: mmurphy@swin.edu.au}, Wim Ubachs$^{3}$\thanks{E-mail: w.m.g.ubachs@vu.nl}, John K. Webb$^{1}$\thanks{E-mail: jkw@phys.unsw.edu.au}.\\
$^{1}$School of Physics, University of New South Wales, Sydney, NSW, 2052, Australia\\
$^{2}$Centre for Astrophysics and Supercomputing, Swinburne University of Technology, Victoria, 3122, Australia\\
$^{3}$Department of Physics and Astronomy, LaserLaB, VU University, De Boelelaan 1081, 1081 HV Amsterdam, The Netherlands}
\begin{document}

\date{Accepted for publication, 23 June 2011}

\pagerange{\pageref{firstpage}--\pageref{lastpage}} \pubyear{2010}

\maketitle

\label{firstpage}

\begin{abstract}
Molecular hydrogen transitions in quasar spectra can be used to constrain variation in the proton-to-electron mass ratio, $\mu\equiv m_p/m_e$, at high redshifts ($z\gtrsim 2$). We present here an analysis of a new spectrum of the quasar Q0528$-$250, obtained on VLT/UVES (the Ultraviolet and Visual Echelle Spectrograph, on the Very Large Telescope), and analyse the well-known H$_2$ absorber at $z=2.811$ in this spectrum. For the first time we detect HD (deuterated molecular hydrogen) in this system with a column density of $\log_{10}(N/\mathrm{cm^{-2}})=13.27 \pm 0.07$; HD is sensitive to variation in $\mu$, and so we include it in our analysis. Using 76 H$_2$ and 7 HD transitions we constrain variation in $\mu$ from the current laboratory value to be $\Delta\mu/\mu = (0.3\pm 3.2_\mathrm{stat} \pm 1.9_\mathrm{sys})\times 10^{-6}$, which is consistent with no cosmological variation in $\mu$, as well as with previous results from other H$_2$/HD absorbers. The main sources of systematic uncertainty relate to accurate wavelength calibration of the spectra and the re-dispersion of multiple telescope exposures onto the one pixel grid.  
\end{abstract}

\begin{keywords}
quasars: absorption lines, cosmology: observations, methods: data analysis
\end{keywords}

\section{Introduction}

The Standard Model of particle physics contains a number of ``fundamental constants'', the values of which are predicted by the theory, and which must be measured by experiment. Two dimensionless constants of particular importance are the fine-structure constant, $\alpha \equiv e^2/(4\pi \epsilon_0 \hbar c)$, and the proton-to-electron mass ratio, $\mu \equiv m_p / m_e$. The former determines the strength of electromagnetism, whereas the latter is a measure of the relative strengths of the strong and electroweak scales. Detection of spatial or temporal variation in a fundamental constant would directly demonstrate the incompleteness of the Standard Model, and might guide efforts to develop post-Standard Model theories.

Quasar absorption lines can be used to constrain variations in $\alpha$ and $\mu$ across most of the observable universe. In particular, the wavelengths of certain transitions are sensitive to changes in the value of $\alpha$ or $\mu$; if $\alpha$ or $\mu$ were different, the relative wavelengths of the spectral lines would differ from those observed in the laboratory, even after correction for the redshift of the absorbing cloud. By comparing measurements of absorption transitions in quasar absorbers to precise laboratory measurements, one is able to constrain changes in $\alpha$ or $\mu$.

\subsection{Molecular hydrogen}

\citet{Thompson:1975} noted that the Lyman and Werner transitions of molecular hydrogen (with $\lambda_0 \lesssim 1150\mathrm{\AA}$) can be used to constrain variations in $\mu$ at $z\gtrsim 2$, where these UV transitions are redshifted into the optical spectrum and therefore can be observed with ground based telescopes. In particular, one can measure the quantity $\Delta\mu/\mu \equiv (\mu_z - \mu_0)/\mu_0$ where $\mu_z$ is the measurement of $\mu$ at a redshift $z$ and $\mu_0$ is the laboratory value of $\mu$.

If $\mu$ were to vary, the wavelengths of the H$_2$ transitions would vary by different amounts. The shift can be quantified as 
\begin{equation}
 \lambda_i = \lambda_i^0\left(1+z\right)\left(1 + K_i\frac{\Delta\mu}{\mu}\right), \label{eq_1}
\end{equation}
where $\lambda_i^0$ is the rest wavelength of the transition, $\lambda_i$ is the observed wavelength and $K_i$ is a `sensitivity coefficient' determining the magnitude and sign of the effect \citep{Varshalovich:93a}. That is,
\begin{equation}
 K_i = \frac{d \ln \lambda_i}{d \ln \mu}.
\end{equation}
 $z$ is the absorption redshift of the cloud (equivalently, the redshift of a transition with no sensitivity to a change in $\mu$). The determination of $z$ is not degenerate with determination of $\Delta\mu/\mu$ provided that two or more transitions with different $K_i$ values are used.

It is worth noting that, if we define a nominal observation redshift for a particular transition as $z_i$, then
\begin{equation}
 \frac{\Delta v_i}{c} \approx \zeta_i \equiv \frac{z_i - z}{1+z} = K_i \frac{\Delta\mu}{\mu}. \label{eq_LBLFM}
\end{equation}
That is, the velocity shift of an individual line is linearly proportional to $\Delta\mu/\mu$. $\zeta_i$ is known as the `reduced redshift'. The redshift $z$ is that of a transition with $K_i=0$. In practice, $z$ must be measured simultaneously with $\zeta_i$ from a set of transitions (e.g. through regression methods) because it is unknown \emph{a priori}.

Recent indications have emerged for a significantly positive $\Delta\mu/\mu$ from the study of two H$_2$ absorbers at $z=2.595$ and $z=3.025$ toward Q0405$-$443 and Q0347$-$383 respectively. \citet{Ivanchik:05a} analysed high resolving power ($R\sim 50,000$) VLT/UVES spectra of these absorbers, and reported a combined value of $\Delta\mu/\mu$ for the two absorbers of $(30.5\pm 7.5)\times 10^{-6}$ and $(16.5\pm 7.4)\times 10^{-6}$ depending on whether H$_2$ laboratory wavelengths were used from \citet{Abgrall:93a} or \citet{Philip:04a} respectively. \citet{Reinhold:06-1} analysed the same two absorbers using improved laboratory wavelength values, and $K_i$ values derived from a more accurate calculation. Using a table of observed line wavelengths from \citet{Ivanchik:05a}, they derived $\Delta\mu/\mu$ values of $(27.8\pm 8.8)\times 10^{-6}$ for the absorber toward Q0405$-$443, $(20.6\pm 7.9) \times 10^{-6}$ for the absorber toward Q0347$-$383, and a combined value of $(24.5\pm 5.9)\times 10^{-6}$. 

\citet{King:08} analysed the same raw spectra of Q0405$-$443 and Q0347$-$383 using a more comprehensive fitting method in which all transitions are fitted simultaneously with the addition of $\Delta\mu/\mu$ as a free parameter in the fit. They used an improved flux extraction and, importantly, an improved wavelength calibration procedure \citep{Murphy:07b} which should significantly reduce systematic errors. They found $\Delta\mu/\mu = (8.2 \pm 7.4) \times 10^{-6}$ and $(10.1 \pm 6.2)\times 10^{-6}$ for the absorbers in Q0347$-$383 and Q0405$-$443 respectively. They also analysed the $z=2.811$ absorber toward Q0528$-$250, providing a constraint of $\Delta\mu/\mu = (-1.4\pm 3.9) \times 10^{-6}$. A weighted mean of these three values yielded $\Delta\mu/\mu = (2.6\pm 3.0)\times 10^{-6}$.

\citet{Wendt:08a} and \citet{Thompson:09a} also analysed the absorbers toward Q0405$-$443 and Q0347$-$383 using the same UVES spectra, and derived results statistically consistent with $\Delta\mu/\mu = 0$. \citet{Wendt:11a} re-analysed the existing Q0347$-$383 exposures combined with exposures under program ID 68.B-0115(A) that had not been previously analysed, again deriving a result statistically consistent with zero. Given the results of \citet{King:08}, \citet{Wendt:08a}, \citet{Thompson:09a} and \citet{Wendt:11a} it does not appear that there is significant evidence for $\Delta\mu/\mu \neq 0$ from the absorbers in Q0405$-$443 and Q0347$-$383.

\citet{Malec:10} reported a strong constraint from the $z=2.059$ absorber toward J2123$-$0050 from spectra obtained using Keck/HIRES. A combination of good seeing conditions and the brightness of the quasar ($r$-band magnitude $\approx 16.5$ mag) meant that they were able to obtain a $R\sim 110,000$ spectrum of sufficiently high SNR in a single night. They found that $\Delta\mu/\mu = (5.6\pm 5.5_\mathrm{stat} \pm 2.9_\mathrm{sys})\times 10^{-6}$ from 86 H$_2$ transitions. Recently the same J2123$-$0050 absorber system was observed with VLT/UVES at a somewhat lower resolution ($R\sim 53,000$) but with a higher signal-to-noise ratio \citep{vanWeerdenburg:11a}. The spectrum covered 96 H$_2$ and HD lines and yields a result of $\Delta\mu/\mu = (8.5 \pm 3.6_\mathrm{stat} \pm 2.2_\mathrm{sys})\times 10^{-6}$, in very good agreement with the results from Keck.

From consideration of the results above, it does not appear at present that there is any strong evidence for $\Delta\mu/\mu \neq 0$ at $z>1$ from measurements of H$_2$ absorbers. 

Unfortunately the number of quasars known to contain H$_2$ absorption features in their spectra is low ($\lesssim 15)$, which makes obtaining a statistical sample of $\Delta\mu/\mu$ measures from the H$_2$ method difficult. To obtain a reliable measurement of $\Delta\mu/\mu$ from H$_2$, one needs a H$_2$ column density that is high enough to produce significant absorption, but preferably not saturated (which makes determination of line centroids difficult). Similarly, transitions of significantly different $K_i$ must be used to achieve good precision on the measurement of $\Delta\mu/\mu$. Unfortunately, the only quasar absorbers to date which have been able to yield measurements of $\Delta\mu/\mu$ at the $\lesssim 10^{-5}$ level are those listed above, which explains the significant number of analyses on these few objects. 

\subsection{Historical Q0528$-$250 constraints}

Since the absorber in Q0528$-$250 is the focus of this work, we note here previous constraints on variation in $\mu$ from this object.

\citet{Varshalovich:93a} analysed the $R\sim 4,000$ spectrum of \citet{Foltz:88a} of Q0528$-$250 to obtain $|\Delta\mu/\mu|<0.005$. \citet{Varshalovich:95a} reanalysed the same spectrum to conclude that $|\Delta\mu/\mu|<0.002$. \citet{Potekhin:98a} used new observations of the same system at higher resolving power ($R\sim14,000$) to obtain $\Delta\mu/\mu=(-10\pm8)\times10^{-5}$.

\citet{Cowie:Songaila:1995} used a $R=36,000$ Keck observation of Q0528$-$250 to produce $\Delta\mu/\mu\in[-7,5.5]\times10^{-4}$ (95 percent confidence limits).

\citet{Ubachs:04a} used the improved laboratory H$_2$ wavelengths of \citet{Philip:04a} to re-investigate the reported line positions for transitions in the spectra of Q0528$-$250, Q0347$-$383 and Q1232+082. For the combined data, they found that $\Delta\mu/\mu=(-0.5\pm1.8)\times10^{-5}$. Omitting the Q0528$-$250 data, which was of poorer quality, they obtained $\Delta\mu/\mu=(1.9\pm1.5)\times10^{-5}$. An explicit value for Q0528$-$250 was not given.

\subsection{Other quasar constraints on $\Delta\mu/\mu$}

The inversion transitions of ammonia are strongly sensitive to variation in $\mu$, with $K_i\sim 4.2$ \citep{Flambaum:07a}. \citet{Murphy:Flambaum:08} and \citet{Henkel:09} compared the inversion transitions of NH$_3$ with rotational molecules to obtain stringent constraints on $\Delta\mu/\mu$ at $z<1$. \citet{Murphy:Flambaum:08} used B0218+357 to obtain $\Delta\mu/\mu = (0.74\pm 0.47_\mathrm{stat}\pm0.76_\mathrm{sys})\times 10^{-6}$ at $z=0.69$, whilst \citet{Henkel:09} used PKS1830$-$211 to obtain $\Delta\mu/\mu=(0.08\pm 0.47_\mathrm{sys})\times 10^{-6}$ at $z=0.89$. \citet{Kanekar:11a} analysed the $z=0.69$ absorber toward B0218+357 to obtain $\Delta\mu/\mu = (-3.5 \pm 1.2) \times 10^{-7}$.

Constraints derived from the analysis of H$_2$/HD absorbers are not statistically competitive with the ammonia results at present. However, the ammonia results are from lower redshifts ($z<1$), whilst the H$_2$/HD results all probe $z>2$, and therefore probe much larger distances and lookback times. The two methods are thus complementary. 

Recently it has been suggested that H$_3$O$^+$ ions \citep{Kozlov:11a} and CH$_3$OH molecules \citep{Jansen:11a} are even more sensitive probes to detect or constrain $\mu$ variation; these transitions exhibit large sensitivity coefficients ($K_i > 10$).

\subsection{Motivation for re-analysis of Q0528$-$250}

The constraint on $\Delta\mu/\mu$ derived from the Q0528$-$250 absorber presented in \citet{King:08} is very precise. However, there are good reasons to revisit this absorber. Some of the exposures which contributed to the spectrum used in that analysis are not well-calibrated. By well-calibrated, we mean that the exposures do not have ThAr (thorium-argon) calibration spectra taken immediately afterwards. The design of VLT/UVES is such that the position of the spectrograph grating is reset between different quasar exposures. Although the specification is such that the placement of the grating should be good to within 0.1 pixels \citep{dOdorico:00a}, the use of spectra for which the ThAr spectra were taken after grating resets necessarily introduces wavelength calibration uncertainties into the spectrum. 

A more subtle concern for the earlier observations (under program IDs P66.A-0594, P68.A-0600 and P68.A-0106) is that they suffer from the fact that the slit width used for the observations was often significantly larger than the prevailing seeing conditions. For those observations, a 1 arcsecond slit was used in both the blue and red arms of UVES. The average ratio of the slit width to seeing, where seeing is quantified by the output of the DIMM (differential image motion monitor) and the seeing values were weighted by the duration of the exposure, was 0.78. In one exposure of 1.6 hours, the ratio was as low as 0.48. In the case where the seeing is significantly smaller than the slit width, the instrumental profile will be non-Gaussian, which complicates the analysis. Although in principle one can use a numerically-provided instrumental profile to convolve the Voigt profile model with, assuming that the instrumental profile is Gaussian is significantly easier.

\section{Data}

\subsection{Description of the new Q0528$-$250 spectrum}

Note that this paper uses only new spectra of Q0528$-$250 so as to measure $\Delta\mu/\mu$ as independently as possible from that presented in \citet{King:08}.

Our new observations are summarized in Table \ref{tab:obs}. We observed Q0528$-$250 in late 2008/early 2009 under ESO program ID 082.A-0087, with exposures totalling approximately 8.1 hours. All science exposures were followed immediately by ThAr calibration exposures without any intervening grating resets, thereby reducing drifts and shifts in the spectrograph in the intervening time. The average slit-width-to-seeing ratio for these exposures is 1.03; that is, the slit width was more appropriately matched to the average seeing conditions than for the exposures contributing to the spectrum analysed in \citet{King:08}.

We optimally extracted and wavelength calibrated the raw spectra using the ESO UVES Common Pipeline Language (CPL) software suite. While the CPL code redisperses the spectra onto a linear wavelength scale by default, we used only the original, `un-redispersed' flux and error array for each echelle order in subsequent reduction steps. A custom code, {\sc uves\_popler}\footnote{See http://astronomy.swin.edu.au/$\sim$mmurphy/UVES\_popler; maintained by MTM.}, was used to combine these extracted echelle orders from the 10 exposures into a single spectrum. The relevant heliocentric corrections and air--vacuum wavelength conversions were calculated and applied to the wavelength scale of each echelle order before redispersing all orders onto a common, log-linear wavelength scale with dispersion 2.0\,km\,s$^{-1}$\,pixel$^{-1}$. The best-fitting relative flux scaling between overlapping orders from all exposures was determined using an automatic $\chi^2$ minimisation scheme. After applying these relative scalings, the extracted quasar exposures were coadded with inverse-variance weighting and a cosmic ray rejection algorithm. Note that, as implied by equation \ref{eq_1}, effects which shift the wavelength scale by a constant velocity at all wavelengths are generally unimportant\footnote{This is not strictly true when many quasar exposures are combined, as is the case here. If different velocity shifts are applied to the different quasar exposures, and if the relative weights of the exposures vary with wavelength when forming the final, combined spectrum, then small relative velocity shifts will be measured between transitions at different wavelengths.}.

{\sc uves\_popler} was also used to automatically fit a continuum to the final spectrum. This continuum was acceptable for our purposes only redwards of the Lyman-$\alpha$ forest. In the Lyman-$\alpha$ forest the continuum was manually re-fitted with low-order polynomials to obtain a nominal continuum against which the H$_2$ absorption lines of interest could be reliably defined. Local constant or linear modifications to this continuum were necessary in the vicinity of some H$_2$ lines when conducting the detailed spectral fitting, as described below.

Using the CPL pipeline we extracted the ThAr flux with the same spatial profile weights derived for the corresponding quasar exposure, thereby allowing the wavelength calibration polynomial most appropriate to the quasar exposure to be established. The CPL pipeline makes use of the ThAr line-list derived by \citet{Murphy:07b} specifically for accurate calibration of UVES ThAr exposures. The calibration residuals had a root-mean-square (RMS) deviation around the final polynomial of 70\,m\,s$^{-1}$ in UVES's blue arm (where all the H$_2$ lines fall) and 55\,m\,s$^{-1}$ over the majority of the two red arm chips, on average.

From the widths of the extracted ThAr features we derived an average full-width-at-half-maximum (FWHM) resolution of $\approx$5.45\,km\,s$^{-1}$, as expected for a slit width of 0\farcs8 \citep{dOdorico:00a}. We initially used this FWHM value as the instrumental resolution in our spectral fits. However, we expect -- and later demonstrate -- that a somewhat lower value of 5.15\,km\,s$^{-1}$ better represents narrow, unresolved lines in the spectrum because the quasar illuminates the slit centrally rather than uniformly like the ThAr lamp.

\begin{table*}
\begin{center}
  \caption{Journal of VLT/UVES observations of quasar Q0258$-$250. Each row specifies the starting time of a single quasar exposure (column 1), all of which were taken with the DIC1 dichroic splitter, giving a central wavelength ($\lambda_\mathrm{c}$) in the blue (``B'') and red (``R'') arms as shown in column 2. The third column gives the exposure times for the quasar and subsequent ThAr wavelength calibration exposures in each arm. Column 4 gives the slit width used for each arm. These should be compared with the average FWHM seeing, as reported by the on-site Differential Image Motion Monitor (DIMM). The average difference between the spectrograph air temperatures at the times of the quasar and ThAr exposures, $\Delta T\equiv T_\mathrm{QSO}-T_\mathrm{ThAr}$, is given in the sixth column. The final column shows the atmospheric pressure difference between the quasar and ThAr exposures, $\Delta P\equiv P_\mathrm{QSO}-P_\mathrm{ThAr}$. On-chip binning by a factor of 2 in both spatial and spectral directions was used for all exposures.}
\vspace{-0.5em}
\label{tab:obs}
\begin{tabular}{lccccccccccc}\hline
UT date \& time    & \multicolumn{2}{c}{$\lambda_\mathrm{c}$} & \multicolumn{3}{c}{$T_\mathrm{exp}$ [s]} & \multicolumn{2}{c}{Slit-width}   & DIMM seeing  & \multicolumn{2}{c}{$\Delta T$} & $\Delta P$ \\
{[yyyy-mm-dd hh:mm]} & \multicolumn{2}{c}{[nm]}             & QSO & \multicolumn{2}{c}{ThAr}      & \multicolumn{2}{c}{[arcseconds]} & [arcseconds] & \multicolumn{2}{c}{[K]}        & [hPa]      \\
                   & B & R                                &     &  B  & R                       &  B & R                           &              & B & R                          &            \\ \hline
2008-11-23 03:58   & 390 & 564                           & 2900.0 & 1.6 & 1.8 & 0.8 & 0.7 & 0.75 & 0.2 & 0.0 & 0.2 \\
2008-11-23 04:52   & 390 & 580                           & 2900.0 & 1.6 & 1.8 & 0.8 & 0.7 & 0.79 & 0.0 & 0.0 & 0.2 \\
2008-11-25 07:35   & 390 & 580                           & 2900.0 & 1.6 & 1.8 & 0.8 & 0.7 & 0.43 & 0.0 & 0.0 & 0.0 \\
2008-12-23 05:03   & 390 & 580                           & 2900.0 & 1.6 & 1.8 & 0.8 & 0.7 & 0.75 & 0.0 & 0.0 & 0.0 \\
2008-12-23 05:54   & 390 & 580                           & 2900.0 & 1.6 & 1.8 & 0.8 & 0.7 & 0.74 & 0.0 & 0.0 & 0.1 \\
2008-12-23 06:46   & 390 & 580                           & 2900.0 & 1.6 & 1.8 & 0.8 & 0.7 & 0.99 & 0.0 & 0.0 & 0.2 \\
2008-12-23 07:38   & 390 & 580                           & 2900.0 & 1.6 & 1.8 & 0.8 & 0.7 & 1.07 & 0.1 & 0.0 & 0.1 \\
2009-01-25 02:40   & 390 & 580                           & 2900.0 & 1.6 & 1.8 & 0.8 & 0.7 & 0.72 & 0.0 & 0.1 & 0.0 \\
2009-01-26 04:02   & 390 & 564                           & 2900.0 & 1.6 & 1.8 & 0.8 & 0.7 & 1.19 & 0.0 & 0.2 & 0.1 \\
2009-02-26 02:52   & 390 & 580                           & 2900.0 & 1.6 & 1.8 & 0.8 & 0.7 & 0.75 & 0.0 & 0.0 & 0.0 \\\hline
\end{tabular}
\end{center}
\end{table*}

\subsection{Laboratory wavelengths and K$_i$ values}

Until recently, the uncertainties in the laboratory wavelengths for the Lyman and Werner series H$_2$ transitions were comparable to the typical uncertainties in the measured line positions in the quasar spectra. However, a significant amount of recent work has improved this situation by using laser-based spectroscopic techniques to obtain much more precise laboratory values. In particular, \citet{Philip:04a}, \citet{Ubachs:04a} and \citet{Ivanov:08b} improved the fractional wavelength accuracy ($\delta \lambda/\lambda$) to $\sim 5\times 10^{-8}$, whilst \citet{Salumbides:08a} improved the fractional accuracy to $\sim 5\times 10^{-9}$ for most Lyman transitions and $\sim 10^{-8}$ for Werner transitions. See also \citet{Bailly:09a}. 

The $K_i$ coefficients can be calculated either through semi-empirical methods \citep{Ubachs:07a} or from \emph{ab initio} methods \citep{Meshkov:06a}. The $K_i$ values derived from these two processes are in good agreement (to within 1 percent). 

\citet{Malec:10} (see their Table 1) have collated the laboratory wavelength data from the above sources and the $K_i$ values from \citet{Ubachs:07a} to produce a comprehensive list which may be used for an analysis of $\Delta\mu/\mu$. They have used oscillator strengths from \citet{Abgrall:94a} and calculated damping coefficients from \citet{Abgrall:00a}. We therefore have used the laboratory wavelength data, $K_i$ values, oscillator strengths and damping coefficients tabulated in \citet{Malec:10} for our analysis of the H$_2$ absorber in Q0528$-$250.

\subsubsection{HD}

We have detected HD (deuterated molecular hydrogen) at $z \sim 2.811$ in the Q0528$-$250 spectrum with a column density of $\log_{10}(N/\mathrm{cm}^{-2})=13.27\pm0.08$. \citet{Malec:10} have collated oscillator strengths, laboratory wavelength values and $K_i$ values for HD; wavelength values are from \citet{Hollenstein:06-1} and \citet{Ivanov:08a}, $K_{i}$ values are from \citet{Ivanov:08a,Ivanov:10a} and oscillator strengths were calculated by \citet{Malec:10} from Einstein $A$ coefficients given in \citet{Abgrall:06a}. We describe our treatment of HD further in section \ref{s_fitting_HD}.

\section{Methods and methodology}

To estimate $\Delta\mu/\mu$ in Q0528$-$250, we model the H$_2$/HD transitions in the absorber with Voigt profiles using the non-linear least-squares $\chi^2$ minimisation program \textsc{vpfit}\footnote{http://www.ast.cam.ac.uk/$\sim$rfc/vpfit.html}. \textsc{vpfit} was written specifically to fit Voigt profiles to quasar absorption spectra. Each Voigt profile is described by three parameters: the redshift of the transition, $z$, the column density, $N$, and the Doppler width, $b$. Knowledge is also required of the rest wavelength, oscillator strength ($f$) and damping constant ($\Gamma$) for each transition. The total model is the sum of the optical depths of a series of Voigt profiles convolved with a model for the instrumental profile (assumed to be Gaussian).

If each transition in the absorber was well-represented by a single, unblended Voigt profile then determining $\Delta\mu/\mu$ would be straightforward: one would measure a redshift for each H$_2$ transition, and then calculate a set of $\zeta_i$ values (equation \ref{eq_LBLFM}). Equation \ref{eq_LBLFM} implies that a plot of $\zeta_i$ against K$_i$ would have a slope of $\Delta\mu/\mu$. 

In practice, there are two factors which complicate the analysis. Firstly, the H$_2$ transitions all have rest wavelengths less than that of Lyman-$\alpha$ ($1215.7\mathrm{\AA}$), and therefore are found only in the Lyman-$\alpha$ forest, the dense series of broad H\,\iscs transitions found bluewards of the quasar's Lyman-$\alpha$ emission line. This means that the H$_2$ transitions are seen against a background of relatively broad Lyman-$\alpha$ lines found at random redshifts and with varying optical depths. To accurately determine the line centroids of the H$_2$ transitions, one must model the Lyman-$\alpha$ forest transitions simultaneously with the H$_2$ transitions. Similarly, unless one allows the parameters of the Lyman-$\alpha$ transitions to be determined simultaneously with those of the H$_2$ transitions, one will necessarily under-estimate the errors in the line centroids for the H$_2$ transitions and therefore also the error on $\Delta\mu/\mu$. 

Secondly, the Q0528$-$250 absorber is not well-represented by a single Voigt profile. Instead, it shows `velocity structure' -- several clouds closely separated in velocity space but having different optical depths and Doppler widths. Provided that the SNR of the spectrum is sufficiently high, one can decompose the observed absorption profile into different `velocity components' (one for each absorption cloud). A reasonable model of the absorber is necessary if one is to obtain an accurate estimate of $\Delta\mu/\mu$. 

In order to estimate $\Delta\mu/\mu$ for the Q0528$-$250 absorber it is therefore necessary to: \emph{i)} construct a Voigt profile model for the Q0528$-$250 H$_2$ absorber, determining the required number of velocity components using objective criteria, and; \emph{ii)} construct a model for the Lyman-$\alpha$ forest in the vicinity of each of the H$_2$ transitions. These two steps constitute almost all of the effort in accurately measuring $\Delta\mu/\mu$ from an existing spectrum of the absorber (significant work being required to obtain a spectrum). In section \ref{s_model_selection} we discuss the statistical criteria used to determine the preferred model. In section \ref{s_constructing_vp_model} we discuss how we build up a model of the Lyman-$\alpha$ forest in the vicinity of the H$_2$ transitions. In section \ref{s_determining_num_H2_components} we describe how we determine the best-fitting number of H$_2$ velocity components. 

\subsection{Model selection}\label{s_model_selection}

\citet{Webb:99,Webb:01,Webb:11a}, \citet{Murphy:01b,Murphy:03,Murphy:04,Murphy:08} and \citet{King:11a} discussed the importance of accurately constructing a Voigt profile fit to an absorber in the context of potential variation of $\alpha$. We apply a similar methodology here in attempting to create the simplest Voigt profile model which best explains the absorption spectrum. \citet{King:08} and \citet{Malec:10} discussed the importance of using an appropriate number of velocity components when fitting H$_2$ absorbers to estimate $\Delta\mu/\mu$. 

When comparing two models with the same number of free parameters, under the maximum likelihood method one should prefer whichever model has the lowest $\chi^2$. However, when comparing models with different numbers of free parameters, one must consider whether the reduction in $\chi^2$ as a result of the addition of more free parameters is sufficiently large to justify the extra free parameters; a so-called `information criterion' should be used to select the most appropriate model. 

One statistic which can be used to compare models with different numbers of free parameters is the Akaike Information Criterion \citep{Akaike:74}, defined as $\mathrm{AIC} = \chi^2 + 2p$, where $p$ is the number of free parameters. As the AIC is only correct in the limit of large $n/p$ (where $n$ is the number of spectral points included in the fit), we use the AIC corrected for finite sample sizes, \citep{Sigiura:1978}, defined as
\begin{equation}
 \mathrm{AICC} = \chi^2 + 2p + \frac{2p(p+1)}{n-p-1}.
\end{equation}
The number of degrees of freedom, $\nu$, is therefore given by $\nu = n-p$. If several competing models are being considered, one chooses the model which has the lowest AICC. The actual value of the AICC is not important; only relative differences matter. The suggested interpretation scale for the AICC is the Jeffreys' scale \citep{Jeffreys:1961} \citep[see][for further discussion]{Liddle:07}, where $\Delta \mathrm{AICC} = 5$ and $\Delta \mathrm{AICC}=10$ are considered strong and very strong evidence against the weaker model respectively. 

When constructing the Voigt profile model, our goal is to find a physically plausible model for the spectral data which minimises the AICC.

\subsection{Constructing the Voigt profile model}\label{s_constructing_vp_model}

We built up our model of the molecular hydrogen transitions and the surrounding forest through an iterative process which we describe here. With knowledge of the redshift of the molecular hydrogen absorbers, in each spectrum we searched for molecular hydrogen transitions which we considered to be potentially usable. We consider potentially usable transitions to be those for which the molecular hydrogen transition can be visually distinguished from its surrounds. This necessarily precludes the use of H$_{2}$ transitions in regions of near zero flux, but in any event these transitions would contribute no meaningful constraint on $\Delta\mu/\mu$. 

\subsubsection{Model construction process and physical assumptions}

We did not build up our region fits from those used in \citet{King:08}, but rather started afresh in order to provide as independent a measurement of $\Delta\mu/\mu$ as possible. From a list of potentially usable transitions, we then selected a spectral region around the H$_{2}$ transition, where the region should be large enough to include any absorption feature which might overlap with the H$_{2}$ transition. In general, we attempted to ensure that the fitting region was sufficiently large so as to return to the local continuum, although this was not always possible. In each of the fitting regions, we modelled the molecular hydrogen transition and then modelled all surrounding features as H\,\isc. To do this, we added and removed H\,\iscs components to attempt to achieve a statistically satisfactory model, using the criterion set out in section \ref{s_model_selection}. The line parameters ($N$,$b$,$z$) for different H\,\iscs transitions are independent of each other. Note that although most transitions observed in the forest are indeed due to H\,\isc, there are also metal transitions from other absorbers along the line of sight (including galactic and atmospheric lines). The identification of the origin of these transitions is not necessary if they do not overlap with the H$_{2}$ transitions; we simply modelled them as H\,\iscs in order to have a physical model for them. We describe the treatment of metal lines which overlap with H$_{2}$ lines below. For all transitions assumed to be H\,\iscs (which we refer to hereafter as just H\,\iscs transitions), we use only the $\lambda1215.7$ transition rather than the whole Lyman series, to prevent line misidentification spuriously impacting regions blueward of that transition. Where Lyman-$\beta$ transitions exist in the blue region of the spectrum, we simply modelled them with additional Lyman-$\alpha$ components. 

We then combined models from the regions fitted individually into a model where the regions are fitted simultaneously. As the line parameters for the individual H$_{2}$ transitions were independent when the regions were fitted independently, at this stage we imposed physical restrictions on the transitions by tying certain parameters together to reflect the fact that the transitions are physically connected. The ground states of the H$_2$ and HD molecules have several rotational sub-states described by the quantum number $J$. We refer to these sub-states as being different $J$-levels. For H$_2$ and HD each $J$-level has a different ground state population. This means that the column density, $N$, should be the same for each velocity component arising from a particular $J$-level. The same applies for the Doppler width, $b$, and the redshift, $z$. We impose the assumption that the velocity structure is the same in all the $J$-levels, that is that a velocity component has the same value of $z$ in all $J$-levels. We describe later the impact of this assumption. 

\citet{King:08} did not actually impose the requirement that each velocity component arising from the same $J$-level has the same $N$; instead they fitted the column densities for each transition as free parameters. We apply a similar approach. We must, however, ensure that a physical consistency is maintained, in that the ratios of the line strengths between different velocity components should be the same for transitions arising from the same $J$-level. We therefore imposed the requirement that the ratio of the column densities between the different components was the same for transitions arising from the same $J$-level. In this way, the total column density (effectively, oscillator strength) for each transition was a free parameter, but the ratios of the individual column densities of different velocity components within each transition were constrained. 

To model the forest, we thus iteratively refined the fit by alternately allowing \textsc{vpfit} to minimise $\chi^{2}$ for a particular model, then attempted to improve that model through the addition and deletion of H\,\iscs components to obtain a robust model according to the criteria in section \ref{s_model_selection}.

\subsubsection{Blending with metal lines}

During the iterative process, it can become clear that a molecular transition is blended with another line (presumed H\,\isc) when it was not thought to be from a fit to just that region. This is because the information from the other molecular hydrogen transitions imposes a strong constraint on the $b$ parameters and redshifts of that transition, thus uncovering apparently hidden blends. These blends necessitate the addition of H\,\iscs components that overlap with the H$_{2}$ or HD transition in question. With the addition of extra H\,\iscs transitions, an acceptable fit can generally be achieved. This demonstrates the utility of fitting all transitions simultaneously: otherwise-inconspicuous blends are generally revealed. 

In a few instances, the transitions which had to be included to achieve a statistically acceptable fit had extremely narrow $b$ parameters ($b\lesssim5\,\mathrm{km\, s^{-1}}$). In this case, it is likely that the blend is a metal line from an unknown absorber along the line of sight. As a result, we excluded such H$_2$ or HD transitions from our fit. The reason for not accepting transitions affected by narrow-$b$ interlopers is that any inaccuracy in modelling the interloping transitions could lead to a significant bias in measuring the H$_{2}$ line position -- the narrow $b$ parameter(s) of the interloping transition(s) means that the absorption they cause varies rapidly across the H$_{2}$ line profile. Ultimately, the joint fit of all the molecular hydrogen transitions allows the detection and rejection of transitions which are likely to be contaminated by metal lines. 

Rejecting transitions which are suspected to be contaminated cannot bias $\Delta\mu/\mu$ away from zero. Moreover, this should not bias $\Delta\mu/\mu$ significantly. If the suspicion of contamination in particular lines was in fact due to $\Delta\mu/\mu\neq0$, we would expect to see this problem more frequently, and more obviously, for transitions with larger $\lvert K_{i} \rvert$. The number of transitions rejected was small, and did not appear to be correlated with $\lvert K_{i} \rvert$, and hence it is unlikely that we are biasing $\Delta\mu/\mu$ towards zero.

\subsubsection{Continuum and zero level fitting}

Determination of the local quasar continuum is difficult in the Lyman-$\alpha$ forest due to absorption which can cover many Angstroms. \citet{Malec:10} discussed the fact that different investigators may manually fit qualitatively similar but quantitatively different continuua to particular regions of a spectrum. For this reason, in any region where it appears that the local continuum has not been well determined, we allow for a freely floating continuum that is described by either a constant offset or a linear function, depending on the region in question. The use of a continuum fit which is at most linear helps to prevent strong degeneracies with the parameters of the Voigt profiles which might occur if a higher degree polynomial was used. The parameters of the continuum fit are determined simultaneously with all other parameters in the fit, and therefore any uncertainty in determining the continuum level naturally flows into the uncertainty on $\Delta\mu/\mu$.

Although weak night sky emission is subtracted as part of the initial flux extraction, the subtraction process does not appear to be optimal in the UVES \textsc{cpl} pipeline. In particular, in the base of saturated lines there appears to be a residual flux of about 2 percent of the local continuum. Therefore, in any region which includes saturated lines we allow for the zero level to be a free parameter in the fit, to be determined simultaneously with all other parameters. 

\subsubsection{Over-fitting}

It is possible to add too many H\,\iscs components to a particular region, leading to ``over-fitting''. Over-fitting is undesirable for several reasons. The primary reason is that it means that another, simpler model can explain the absorption spectrum better than the over-fitted model. Perhaps more importantly, it means that the performance of the optimisation algorithm can be substantially impaired. With significant over-fitting, convergence to the $\chi^{2}$ minimum can be excessively slow. In extreme cases, convergence may not occur at all. Over-fitting can be detected through two means:

\emph{i)} The addition of components which increase the AICC suggests that the components are not supported by the data. If the AICC significantly decreases upon removal of the components, this suggests that the model was over-fitted.

\emph{ii)} Over-fitting causes the uncertainty estimates on the parameters of the components in question to be excessively large \citep{GMW:86}. In fact, this is often a good way to directly identify components which are potentially unnecessary; the AICC relates to the model as a whole and therefore cannot suggest which components may be unnecessary. In particular, H\,\iscs transitions with $\sigma_{\log_{10}N}\gtrsim1.0$ or $\sigma_{b}/b\gtrsim1$ are certainly suspicious. In regions with substantial over-fitting, errors can easily be substantially larger than this. The numerical cause of these large errors is strong relative degeneracies between parameters. That is, $\chi^{2}$ is almost flat in some direction in the parameter space relating to the offending transitions. It is this flatness in $\chi^{2}$ which is the cause of poor convergence. Nevertheless, the presence of large errors on some components does not mean that they are unnecessary. In particular, the column densities for transitions which are saturated can be very poorly determined. This necessarily means that saturated H\,\iscs transitions will have large errors on the column density. 

Because of the impact of over-fitting on the convergence of \textsc{vpfit}, we spent considerable effort trying to identify cases of over-fitting, and removed H\,\iscs components as necessary to minimise the problem. 

\subsubsection{Determination of the number of velocity components in the Q0528$-$250 absorber}\label{s_determining_num_H2_components}

\citet{King:08} noted that at least 2 velocity components are plainly required to achieve an adequate fit to the Q0528$-$250 absorber. They argued on the basis of the AICC that 4 velocity components were required. We followed a similar procedure here, by comparing models with two, three and four velocity components. We choose whichever model has the lowest AICC (see section \ref{s_velstruc+results} below).

\subsection{Voigt profile fits}

Our final fits were obtained where we were not able to obtain any statistically appreciable improvement. We show one part of the H$_2$/HD and Lyman-$\alpha$ Voigt profile fit in Fig.\   \ref{fig_example_VP_fit}; the full model may be found in the online-only version in Appendix \ref{app_voigt_profile_fits}. 

\begin{figure*}
\includegraphics[bb=62 181 544 801,width=160mm]{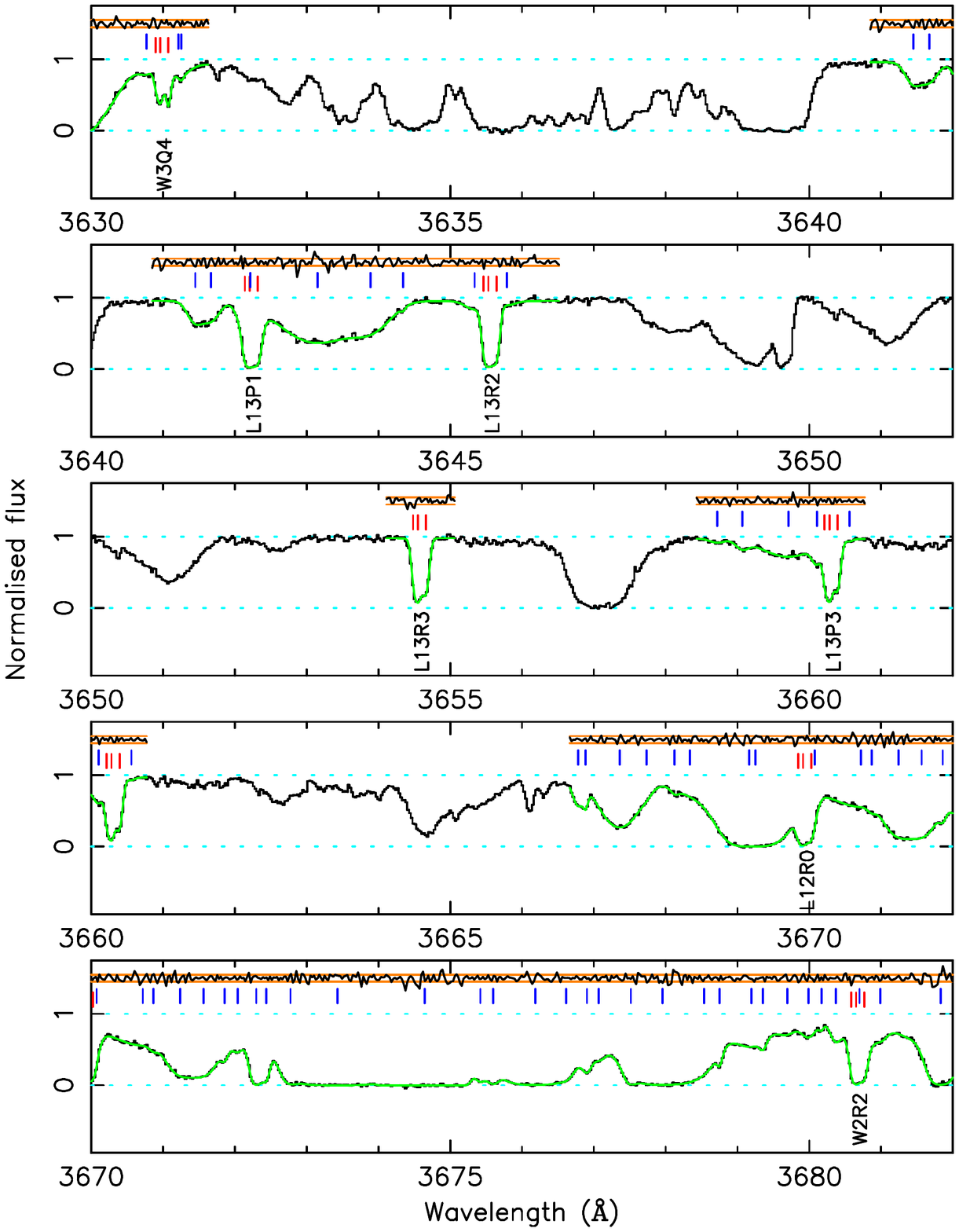}
  \caption{Part of the H$_2$/HD Voigt profile fit for the $z=2.811$ absorber toward Q0528$-$250. The vertical axis shows normalised flux. The model fitted to the spectrum is shown in green. Red tick marks indicate the position of H$_2$/HD components, whilst blue tick marks indicate the position of blending transitions (presumed to be Lyman-$\alpha$). Normalised residuals (i.e. [data - model]/error) are plotted above the spectrum between the orange bands, which represent $\pm 1\sigma$. Labels for the H$_2$ transitions are plotted below the data. Other regions shown in Appendix \ref{app_voigt_profile_fits} (online only).\label{fig_example_VP_fit} }
\end{figure*}

\subsection{Determining $\Delta\mu/\mu$}

There are two methods in the literature used for determining $\Delta\mu/\mu$ for a particular H$_2$ absorber. These can be described as the `line-by-line fitting method' (LBLFM) and the `comprehensive fitting method' (CFM).

In the LBLFM, for each transition one calculates the $\zeta_i$ values in equation \ref{eq_LBLFM}. Equation \ref{eq_LBLFM} then implies that a plot of $\zeta_i$ vs $K_i$ will have a slope of $\Delta\mu/\mu$. The slope of a linear fit to $\zeta_i$ vs $K_i$ can easily be obtained from standard $\chi^2$ minimisation techniques. This method has the advantage of assisting visual detection of problematic transitions, where $\zeta_i$ deviates from the general trend shown by other lines. A strong deviance of this nature implies that a problem exists in the fit to the transition which generated that $\zeta_i$ value; one can then return to the spectral fit to attempt to determine the problem. This method was used by \citet{Ivanchik:05a} and \citet{Reinhold:06-1}.

Unfortunately, the LBLFM cannot be used when the absorber contains two or more closely spaced velocity components. The If velocity components overlap then the calculated errors in the line centroids for those velocity components will be correlated. It follows that, to use the LBLFM, one must first demonstrate that a single component is the best model for the H$_2$/HD lines i.e. that a two component model is statistically worse than a one component model. The LBLFM can be generalised to fit closely spaced velocity components through the use of generalised least squares methods, which allow for correlations between data points. However, to the best of our knowledge this has not been used in the literature to determine $\Delta\mu/\mu$ from a quasar spectrum.

The CFM allows one to reliably measure $\Delta\mu/\mu$ in the case where the absorber contains two or more closely spaced velocity components. In the CFM, one fits all H$_2$ transitions simultaneously with a single redshift for each velocity component, and allows for $\Delta\mu/\mu$ as a free parameter in the fit, such that the rest wavelengths are perturbed as $\lambda_i^0 \rightarrow \lambda_i^0[1+K_i (\Delta\mu/\mu)]$. The value of $\Delta\mu/\mu$ at the $\chi^2$ minimum is the best estimate of $\Delta\mu/\mu$ from the spectrum. In this way, the overlap of the velocity components is naturally accounted for within the fitting process, including the error estimate for $\Delta\mu/\mu$. Additionally, this method significantly reduces the number of free parameters in the fit, which should improve the reliability of the method. This method was used by \citet{King:08} and \citet{Malec:10}.

We have assumed that the errors in the $K_i$ values (which are $\approx 1$ percent) are negligible. This assumption can be clearly justified by examining typical reduced redshift plots e.g. Fig.\ 1 in \citep{King:08}, where it is clear that the uncertainty in determining the velocity shifts of the line centroids is much more important than $\approx 1$ percent errors in the $K_i$ coefficients.

The absorber in Q0528$-$250 contains several closely spaced molecular velocity components. We therefore use the CFM to derive our estimate of $\Delta\mu/\mu$. We note that our fits to each region were all constructed assuming $\Delta\mu/\mu = 0$; $\Delta\mu/\mu$ was allowed to vary only once our model for the H$_2$/HD and Lyman-$\alpha$ lines was complete. If $\Delta\mu/\mu$ is significantly different from zero, it is conceivable that this method could bias our measurement of $\Delta\mu/\mu$ towards zero by ``fitting away'' any $\mu$ variation through the choice of the Lyman-$\alpha$ forest model. However, because the forest lines generally have high $b$ parameters and because our model cannot fit the spectra in a totally arbitrary fashion (our model consists of a series of Voigt profiles, which have rather less flexibility to fit the spectrum than polynomials, for example) any bias induced should be small. Note that, on account of the above methodology, we cannot bias $\Delta\mu/\mu$ away from zero.

\subsection{HD}\label{s_fitting_HD}

HD is sensitive to a change in $\mu$, and therefore here we include HD in our analysis of $\Delta\mu/\mu$. Although HD should display a similar velocity structure to H$_{2}$, the low optical depth of the HD transitions and the small number of transitions observed means that any such structure is unresolved. We therefore model the HD absorption with only a single velocity component. That is, the constraint on $\Delta\mu/\mu$ from HD is derived only by considering potential velocity shifts of the HD lines with respect to each other, and not with respect to H$_2$ transitions. The small number of HD transitions used means that the statistical constraint on $\Delta\mu/\mu$ derived from HD is weak compared to that derived from the H$_2$ transitions. However, future detected HD absorbers observed with greater optical depths and higher SNRs may provide statistically competitive constraints. 

\subsection{Spectral extraction problems}

The \textsc{cpl} pipeline appears to incorrectly estimate the errors associated with the flux data points in the base of saturated lines. In particular, the dispersion of the flux data points about the local zero level is too large to be accounted for by the statistical error. The statistical errors on the flux points appear to be under-estimated by a factor of at least 2. Although it is difficult to determine precisely what happens in regions of low, but non-zero flux, we believe that the errors there are also underestimated. The effect of this is to give falsely high precision on any quantity derived from these data points (including $\Delta\mu/\mu$). Additionally, one cannot fit plausible models to data involving regions of low or negligible flux. 

When we combine the individual exposures into a final spectrum using \textsc{uves\_popler}, the program calculates a check on the consistency of the exposures contributing to each pixel. As the combined value of each flux pixel is given as a weighted mean of pixels from the contributing spectra, this consistency check is a value of $\chi^2_\nu$ for each pixel about the weighted mean. We attempt to correct for the under-estimation of uncertainties in regions of low or negligible flux by applying the following algorithm. For each pixel in the combined spectrum, we take a region of five pixels centred on that point. We then take the median of the $\chi^2_\nu$ values just described that are associated with those five points. We then multiply the error estimate for that spectral pixel by the square root of that median value (that is, $\sigma_i \rightarrow \sigma_i \times \sqrt{\mathrm{median}[\chi^2_\nu]}$). 

However, we only increase the error array if $\mathrm{median}(\chi^2_\nu) > 1$. This is because the spectral errors are generated from photon-counting, and therefore the statistical uncertaintiy should be a lower bound on the true uncertainty. Thus, it is difficult to justify decreasing the error array without particular evidence that the spectral uncertainty is systematically under-estimated.

\section{Results}

In section \ref{s_velstruc+results} we describe our determination of the velocity structure of the absorber, and the $\Delta\mu/\mu$ values (and associated statistical errors) which result. In section \ref{s_consistencychecks} we describe various consistency checks we make to ensure that our results are robust. In section \ref{s_systematics}, we detail our analysis of potential systematic errors. In section \ref{s_finalresult} we give our preferred result including the final estimate of systematic errors. 

\subsection{Transitions used}

\begin{figure}
\includegraphics[bb=47 187 544 766,width=82.5mm]{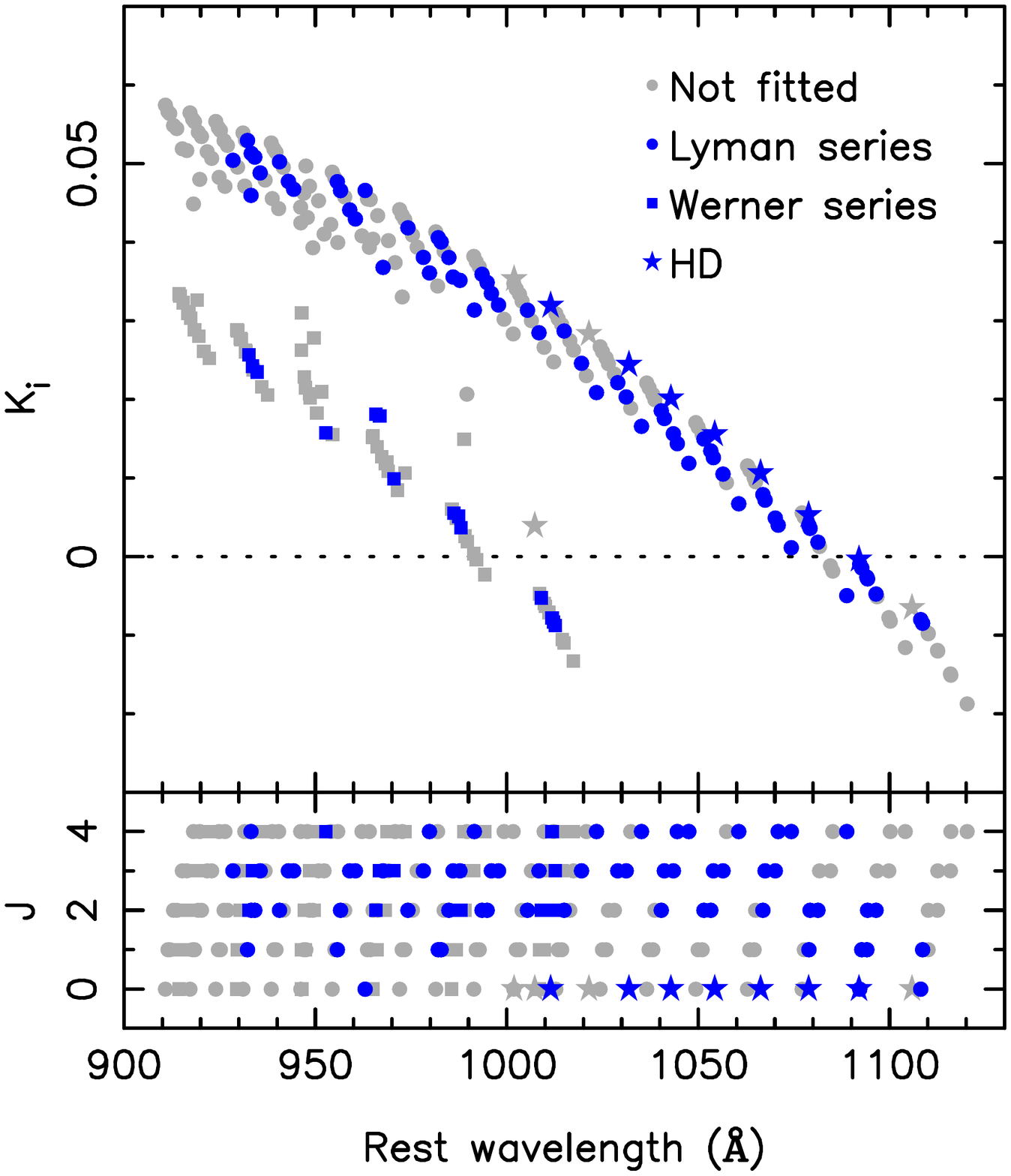}
  \caption{Relationship of $K_i$ and $J$ with $\lambda_0$ (rest wavelength) for the H$_2$ and HD transitions used in our analysis of the $z=2.811$ absorber toward Q0528$-$250. Only H$_2$ transitions with $J\in[0,4]$ are shown. All the HD transitions have $J=0$. \emph{Upper panel}: the sensitivity coefficients, $K_i$, for the transitions used in our analysis of the absorber (dark blue points) and not detected or not fitted (grey points). \emph{Lower panel}: the distribution of transitions with wavelength according to their $J$-level. \label{fig_lambdakiplot}}
\end{figure}

\begin{figure}
\includegraphics[bb=29 153 546 801,width=82.5mm]{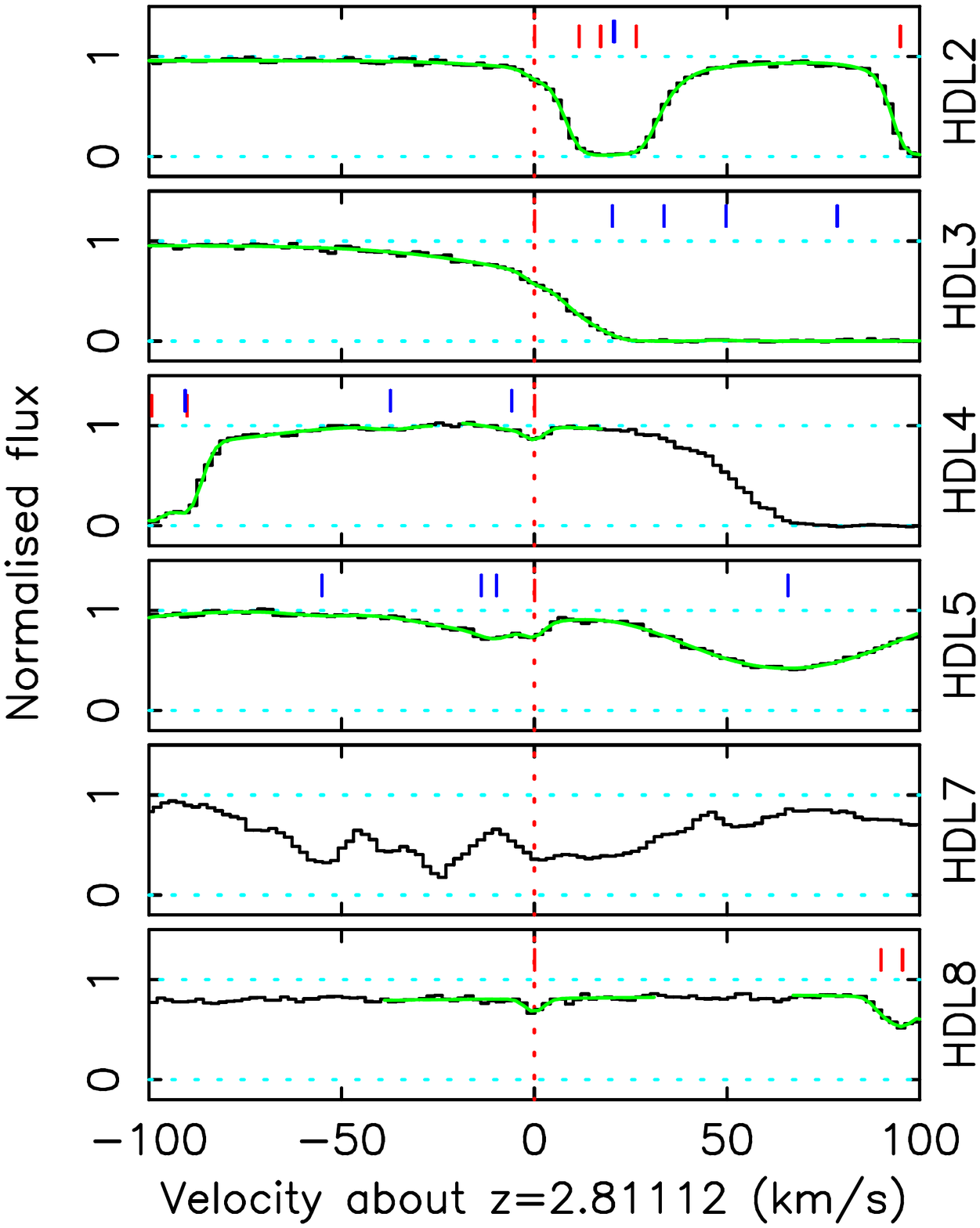}
  \caption{Different spectral regions for our spectrum of Q0528$-$250 demonstrating the presence of HD in the absorber at $z=2.811$. The velocity is shown about $z=2.81112$, which is the fitted HD redshift. The black line is the spectral data and the green line is the model fitted to the data. The red tick marks at $v=0$ are fitted HD transitions. The red tick marks at $v\neq 0$ are fitted H$_2$ components. The blue tick marks are fitted Lyman-$\alpha$ components.\label{fig_HDvelplot}}
\end{figure}

For our analysis, we have included 76 molecular hydrogen transitions. In Fig.\ \ref{fig_lambdakiplot} we give the relationship of $K_{i}$ and $J$ with $\lambda_{0}$ for the H$_2$ and HD transitions used in our analysis. The seven HD transitions we have used are: L1R0, L2R0, L3R0, L4R0, L5R0, L6R0, L8R0. We show the spectral regions for the stronger HD transitions as a velocity plot in Fig.\ \ref{fig_HDvelplot}, clearly demonstrating the existence of HD.

\subsection{Velocity structure and $\Delta\mu/\mu$ results}\label{s_velstruc+results}

We examine the question of the velocity structure of the Q0528$-$280 absorber under different scenarios. To do this, we consider models with two, three and four velocity components. In the 3- and 4-component case, we applied a model in which corresponding velocity components in transitions from the same $J$-level had the same $b$ parameter [$b=F(J)$], and also the scenario in which corresponding transitions had the same $b$ parameter regardless of $J$ [$b\neq F(J)$]. For the 2-component case, we only considered $b=F(J)$. We give the AICC for these scenarios, and the resulting values of $\Delta\mu/\mu$, in Table \ref{tab_numcomponents}. We note that we were unable to obtain a stable fit for a 4-component model where $b$ parameters for corresponding components were forced to be the same for all $J$-levels. In this model, the column density of one of the components in the $J=1$ transitions was driven down to effectively zero. This component was the second strongest component in the $J=2$ and $J=3$ transitions. Although we could have omitted this component, the substantial differences in relative strength between the different components in the different $J$-levels means that this model is very unlikely to be a good representation of the physical situation, and therefore that the value of $\Delta\mu/\mu$ derived is unlikely to be accurate. 

\begin{table}
  \begin{center}
  \caption{Analysis of the velocity structure of Q0528$-$250. $n$ gives the number of components. The second column defines whether the $b$-parameter for different components is fixed or is different for transitions with different $J$, i.e. whether $b$ is a function of $J$ [$b=F(J)$]. The column labelled AICC gives the AICC for the fit. The column $\Delta \mathrm{AICC}$ shows the difference of the AICC with respect to the best-fitting model. For the 4-component, $b\neq F(J)$ model, the column density of one component of the $J=1$ transitions was driven to zero, and the component was rejected, where this component was strongly detected in other $J$-levels. This implies that the model is not physically realistic, and so we label it as unstable. For these fits, the instrumental resolution was assumed to have a FWHM of $5.45\,\mathrm{km\,s^{-1}}$. Note that the AICC is numerically quite large because of the thousands of degrees of freedom due to thousands of spectral points being included in the fit. \label{tab_numcomponents}}
  \begin{tabular}{lclrlc}
  \hline
  $n$ & $b \stackrel{?}{=} F(J)$ & AICC & $\Delta \mathrm{AICC}$ & $\chi^2_\nu$ & $\Delta\mu/\mu$ $(10^{-6})$\\
  \hline
  2 & $b=F(J)$     & 11545.8      & 57.6  & 1.126 & $2.6\pm2.8$\\
  3 & $b=F(J)$     & 11488.2      & 0     & 1.115 & $0.2\pm3.2$\\
  3 & $b\neq F(J)$ & 11653.8      & 165.6 & 1.141 & $0.4\pm3.2$\\
  4 & $b=F(J)$     & 11510.4      & 22.2  & 1.117 & $3.4\pm3.7$\\
  4 & $b\neq F(J)$ & Unstable fit & n/a   & n/a   & n/a\\
  \hline
  \end{tabular}
  \end{center}
\end{table}

We note the sign change in $\Delta\mu/\mu$ from the result in \citet{King:08} [$\Delta\mu/\mu = (-1.4 \pm 3.9) \times 10^{-6}$], although in the $n=3$ case the result is only marginally different from zero. It is clear from the $n=3$ results that a model where different $J$-levels have different $b$ parameters is preferred very strongly over a model with the same $b$-parameter for each $J$-level. This accords well with the results of \citet{King:08}. 

We note that, for these fits, the 3-component model is preferred to the 4-component model, at about the same statistical significance as the 4-component model was preferred to the 3-component model in \citet{King:08}. There are several points to note here:

\emph{i)} The seeing conditions for the spectra used by \citet{King:08} were quite variable, and may have induced a significantly non-Gaussian instrumental profile. The requirement for 4 components in \citet{King:08} may be a reflection of the non-Gaussian profile, rather than the absorber itself. Because the components are unresolved at these resolving powers, identification of the correct number of components is difficult. Nevertheless, what is important is that all the statistical structure in the measured profiles is adequately accounted for. This is directly reflected in the AICC.

\emph{ii)} Some of the exposures contributing to the spectrum used by \citet{King:08} were poorly calibrated. It is conceivable that wavelength miscalibrations could cause a model with more complexity to be favoured.

\emph{iii)} Although we have attempted to apply the same forest model in analysing the 3- and 4-component model (where in each case $\chi^{2}$ is obviously minimised with respect to all the parameters), the construction of the forest model itself depends on the choice of the H$_{2}$ model. Strong H$_{2}$ components will obviously have little effect on the forest model because they are clearly distinguished from the forest. However, the 4th component of the model is weak and unresolved visually. We created the 3-component fit by removing the weakest component from the 4-component fit. In most regions, the resulting fit was reasonable, however in a small number of regions we found that we had to add weak forest components to account for the removal of the 4th H$_{2}$ component. To achieve a like-with-like comparison, we included these extra forest components in the 4-component fit. This means that any test for the statistical significance of the number of components depends somewhat on the choice of forest model near the H$_{2}$ lines. 

\emph{iv)} We noted whilst we were iteratively refining the model in the 3- and 4-component cases that the 4-component model was preferred for most of the refining process, with $\Delta\mathrm{AICC\approx10}$ in its favour for much of the time. It was only in the last few rounds of refining the model that the 3-component model became preferred as a result of changes made to a small number of regions. Therefore, the choice of the correct number of components can be sensitive to decisions made about the forest model in a small number of regions.

\emph{v)} If the instrumental resolution used differs significantly from the true instrumental resolution then one might choose an incorrect number of components.

We conclude from this that although the Jeffreys' scale suggests that there is very significant evidence for the 3-component model over the 4-component model, in light of the fact that this evidence is conditioned on the correct choice of forest model and instrumental resolution, the actual preference for the 3-component model over the 4-component is rather weak. These arguments apply similarly to the preference for a 4-component model over a 3-component model in \citet{King:08}. The issue of whether there are 3 or 4 components is simply very difficult to resolve given the actual SNR and resolution of the spectra available. 

\begin{table}
  \begin{center}
  \caption{Effect of varying the instrumental resolution from FWHM=$5.45\,\mathrm{km\,s^{-1}}$ (table \ref{tab_numcomponents}) to FWHM=$5.15\,\mathrm{km\,s^{-1}}$ on estimates of $\Delta\mu/\mu$ for different models. $n$ gives the number of components. The second column defines whether the $b$-parameter for different components is fixed or is different for transitions with different $J$, i.e. whether $b$ is a function of $J$ [$b=F(J)$]. The column labelled AICC gives the AICC for the fit. The column $\Delta \mathrm{AICC}$ shows the difference of the AICC with respect to the best-fitting model. \label{tab_numcomponents_2}}
  \begin{tabular}{lclrlc}
  \hline
  $n$ & $b \stackrel{?}{=} F(J)$ & AICC & $\Delta \mathrm{AICC}$ &  $\Delta\mu/\mu$ $(10^{-6})$\\
  \hline
  3 & $b=F(J)$     & 11470.4      & 0     &  $0.3\pm3.2$\\
  4 & $b=F(J)$     & 11500.1      & 29.8  &  $4.6\pm3.8$\\
  \hline
  \end{tabular}
  \end{center}
\end{table}

To investigate the effect of varying the instrumental resolution, we re-ran the 3- and 4-component, $b=F(J)$ fits with an instrumental resolution of $5.15\,\mathrm{km\,s^{-1}}$. The results of this are given in Table \ref{tab_numcomponents_2}. There are three points to note from varying the instrumental resolution: \emph{i)} in both cases the AICC decreases, suggesting that the lower instrumental resolution is preferred; \emph{ii)} the preference for the 3-component model over the 4-component model increases, and; \emph{iii)} $\Delta\mu/\mu$ is relatively insensitive to the instrumental resolution, as one would naively expect (particularly for the 3-component model). We give the resulting parameters from this fit in Table \ref{tab_parmvalues}.

\begin{table*}
\begin{center}
  \caption{Best-fitting H$_2$/HD model parameters in the 3 velocity component model by $J$-level. $n$ gives the number of H$_2$ transitions fitted for that $J$-level. The columns labeled ``Component 1'' through ``Component 3'' give the parameters for the 3 H$_2$ components, whilst the HD column gives the parameters for the single fitted HD component. For the H$_2$ components, $N_\mathrm{rel}$ specifies the relative strengths of the three components, defined as $N_{\mathrm{rel},i}=N_i / \Sigma_j N_j$ for the components in that $J$-level; because we fit the line intensities as free parameters for the H$_2$ transitions it is difficult to derive an accurate estimate of the column density. For an approximate estimate of the total column density for each $J$-level, see Table \ref{tab_NJlevel}. The HD components were fitted using the actual oscillator strengths, and so we give a direct estimate of the column density for the $J=0$ level of HD explicitly. $b$ gives the $b$ parameter for the component. Uncertainty estimates are derived from the covariance matrix at the purported optimisation solution. For certain poorly-determined $b$ parameters, where $\sigma_b / b \gtrsim 1$ these uncertainty estimates will not yield accurate confidence intervals on the $b$-parameters due to the fact that $b>0$. Our optimiser has a hard lower limit for $b$ of $0.4\,\mathrm{km\,s^{-1}}$ to prevent numerical problems with the convolution of the Voigt profile model with the instrumental profile model, which is why the $b$ parameters for the $J=3$ and $J=4$ levels of component 3 are both $0.40\,\mathrm{km\,s^{-1}}$. Note the trend for decreasing $b$ with increasing $J$ for the H$_2$ components, which is why the $b=F(J)$ models are significantly preferred over the $b\neq F(J)$ models. \label{tab_parmvalues}}
\begin{tabular}{crlcccc}
\hline
$J$-level              & $n$ & Parameter                                & Component 1        & Component 2        & Component 3         &   HD\\
\hline
                       &     & $z$                             & 2.8110056(9)      & 2.8111223(7)       &  2.8109346(11)      &   2.8111200(33) \\
\multirow{2}{*}{$J=0$} & \multirow{2}{*}{3}    & $b\ (\mathrm{km\,s^{-1}})$      & $5.36 \pm 0.36$    & $3.56 \pm 0.22$    & $1.09 \pm 0.94$     &   $0.95 \pm 0.87$ \\
                       &     & $N_\mathrm{rel}$                & 0.68             & 0.25               & 0.08                &   $\log_{10}(N/\mathrm{cm^{-2}})= 13.27 \pm 0.07$\\ 
\multirow{2}{*}{$J=1$} & \multirow{2}{*}{8}    & $b\ (\mathrm{km\,s^{-1}})$      & $4.64 \pm 0.99$    & $3.57 \pm 0.25$    &  $2.44 \pm 0.62$    &  \\
                       &     & $N_\mathrm{rel}$                & 0.69               & 0.22               & 0.09                &  \\ 
\multirow{2}{*}{$J=2$} & \multirow{2}{*}{24}    & $b\ (\mathrm{km\,s^{-1}})$      & $4.63 \pm 0.25$    & $3.05 \pm 0.08$    & $0.42 \pm 0.08$     &  \\
                       &     & $N_\mathrm{rel}$                & 0.18               & 0.07               & 0.75                &  \\ 
\multirow{2}{*}{$J=3$} & \multirow{2}{*}{28}    & $b\ (\mathrm{km\,s^{-1}})$      & $3.78 \pm 0.19$    & $3.08 \pm 0.08$    & $0.40 \pm 0.05$     &  \\
                       &     & $N_\mathrm{rel}$                & 0.14               & 0.07               & 0.78                &  \\ 
\multirow{2}{*}{$J=4$} & \multirow{2}{*}{13}    & $b\ (\mathrm{km\,s^{-1}})$      & $3.58 \pm 0.26$    & $1.88 \pm 0.15$    & $0.40 \pm 0.22$     &  \\
                       &     & $N_\mathrm{rel}$                & 0.42               & 0.46               & 0.12                &   \\ 
\hline
\end{tabular}
\end{center}
\end{table*}

In principle, one can account for the uncertainty in the choice of model by constructing a weighted mean of the $\Delta\mu/\mu$ values from each model, where the weights are given by the penalised likelihood [i.e. $w_i = \exp(-\mathrm{AICC}/2)$]. Because the difference in the AICC between the 3-component and 4-component models is $\approx 30$, however, the suppression of the 4-component model is sufficiently large so that any contribution from this model can be neglected.

Thus, on the basis of the statistical results in Table \ref{tab_numcomponents_2}, we choose $\Delta\mu/\mu=(0.3\pm3.2)\times10^{-6}$ as our preferred statistical result for the analysis of $z=2.811$ absorber toward Q0528$-$250. This is obviously consistent with no change in $\mu$.

\subsection{Consistency checks}\label{s_consistencychecks}

We can relax some of the assumptions made in our analysis of this absorber to explore whether they have a meaningful impact on the result. In particular, we explore here whether the result we obtain is significantly affected by our assumptions about the different $J$-levels. We follow a similar procedure to that used by \citet{Malec:10}. We also verify that our optimisation algorithm is functioning adequately.

\subsubsection{Different $\Delta\mu/\mu$ from different $J$-levels}

\begin{figure}
\includegraphics[bb=50 92 520 799,angle=-90,width=82.5mm]{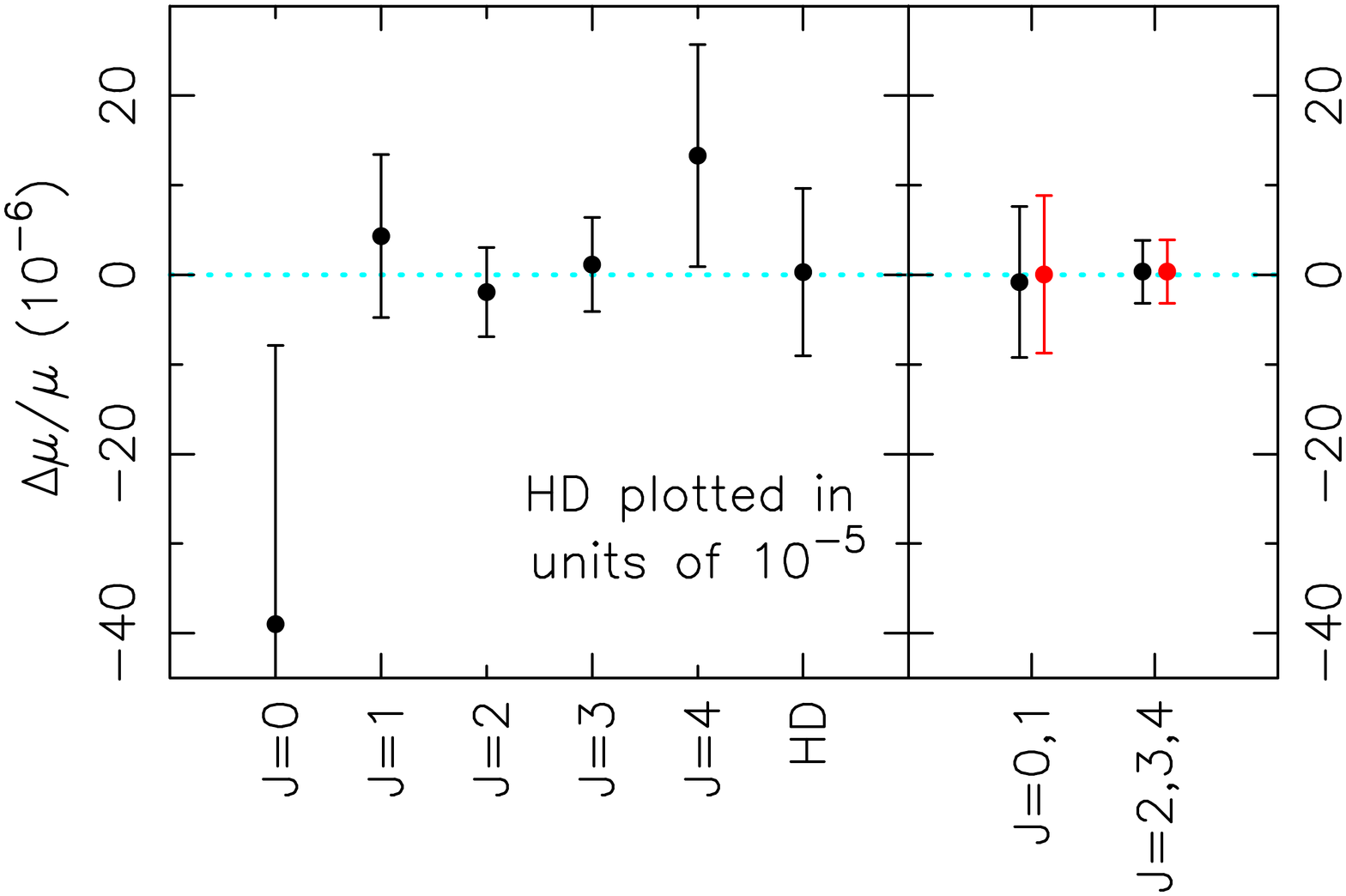}
  \caption{\emph{Left panel:} $\Delta\mu/\mu$ for each H$_2$ $J$-level and HD, assuming that each $J$-level has the same 3-component velocity structure. The $J=0$ level has a substantially larger uncertainty than the $J \in [1,4]$ transitions because only 3 $J=0$ transitions are used in the fit. HD has been plotted in units of $10^{-5}$ to increase clarity for the H$_2$ results. \emph{Right panel:} $\Delta\mu/\mu$ for two groups of transitions, H$_2$ $J\in [0,1]$ (``cold transitions'') and $J\in[2,4]$ (``warm transitions''). The black data points are the results where the redshifts for corresponding components in the cold and warm transitions are assumed to be the same, and the red points are where the redshifts are allowed to differ between the cold and warm transitions. \label{fig_Q0528_diff_J_contrib}}
\end{figure}

Rather than allowing transitions from all $J$-levels to contribute to a single value of $\Delta\mu/\mu$, within \textsc{vpfit} we can calculate a value of $\Delta\mu/\mu$ for each H$_2$ $J$-level (and one for HD) separately. Strictly, the values of $\Delta\mu/\mu$ obtained are not independent because: \emph{i)} they assume that each $J$-level has the same number of velocity components; and, \emph{ii)} the redshifts are tied between corresponding velocity components in transitions arising from different $J$-levels. Nevertheless, this is useful for quantifying the contribution that each $J$-level makes to the final result. \citet{Ubachs:07a} noted that, on account of the para--ortho distribution of H$_{2}$ the $J=1$ state is significantly populated even at low temperatures. They suggested dividing the states into a $J\in[0,1]$ set (cold states) and $J\geq 2$ set (warm states) to examine the impact of temperature. We examine both of these cases in Fig.\   \ref{fig_Q0528_diff_J_contrib}. We see that there is no clear evidence for a difference of $\Delta\mu/\mu$ obtained using transitions arising from different $J$-levels.

\subsubsection{Optimiser performance}

The determinisation of $\Delta\mu/\mu$ requires minimising $\chi^2$ for a model which contains hundreds of free parameters. Adequate performance of the optimiser is crucial in obtaining a reliable value of $\Delta\mu/\mu$ and an associated error. Premature termination of the optimiser will yield an erroneous best estimate for $\Delta\mu/\mu$. Typically this result will be biased towards the value of $\Delta\mu/\mu$ that is used to initalise the optimiser. Code errors or other gross problems with the optimiser will also render the error associated with $\Delta\mu/\mu$ unreliable.

In the context of potential variation of $\alpha$, we have checked that the optimiser within \textsc{vpfit} performs adequately for the case of a few tens of parameters using Markov Chain Monte Carlo (MCMC) methods \citep{King:09}. We found for three systems that the estimated value of $\Delta\alpha/\alpha$ and its associated $1\sigma$ error provided by \textsc{vpfit} are very similar to that determined from MCMC exploration of the likelihood function. Nevertheless, our model for the absorber in Q0528$-$250 and the Lyman-$\alpha$ forest have more than an order of magnitude more free parameters, and so an explicit check on the performance of the optimiser is advantageous. 

A simple check on whether the optimiser is functioning adequately is to restart \textsc{vpfit} with an initial value of $\Delta\mu/\mu$ that is significantly different to both the original starting guess ($\Delta\mu/\mu=0$) and the final result returned by \textsc{vpfit}. To do this, we re-ran the optimisation with a starting value of $\Delta\mu/\mu=10^{-5}$ for the 3-component, $b=F(J)$ fit with an instrumental resolution FWHM of $5.45\,\mathrm{km\,s^{-1}}$. The final value of $\Delta\mu/\mu$ under this circumstance was $0.210\times10^{-6}$, compared to $0.187\times10^{-6}$ when started from $\Delta\mu/\mu=0$. These two numbers differ by $\approx 0.007\sigma$, which is entirely negligible. This demonstrates both that the final result is insensitive to reasonable choices of the starting guess for $\Delta\mu/\mu$ and that our optimiser is functioning adequately.

We note that \citet{Malec:10} performed Monte Carlo simulations on realisations of their J2123$-$0050 spectrum and found consistency between the error on $\Delta\mu/\mu$ and the distribution of $\Delta\mu/\mu$ values. Similar simulations confirm the reliability of \textsc{vpfit} in the context of $\Delta\alpha/\alpha$ \citep{Murphy:PhD,Murphy:03,King:09}

\section{Systematic errors}\label{s_systematics}

\citet{Malec:10} discussed a number of potential systematic errors which could affect the measurement of $\Delta\mu/\mu$ from the analysis of H$_2$/HD absorbers. We investigate the same potential systematic errors here.

It is worth noting initially that a clear potential source of uncertainty arises from the quality of the wavelength calibration. Because the K$_i$ values are well-correlated with wavelength within the Lyman and Werner series (but much less so when considering both the Lyman and Werner series simultaneously), any effect which systematically expands or compresses the wavelength scale will mimic a change in $\mu$. Any effect which only locally perturbs the wavelength calibration will have a random effect on $\Delta\mu/\mu$, and, providing enough transitions are used, the impact of such an effect will average to zero. We discuss wavelength calibration in the next two sections.

\subsection{Known wavelength calibration errors due to uncertainties in the ThAr calibration}

The calibration of the ThAr wavelength scale is not perfect; each of the ThAr transitions displays a residual velocity offset about the best-fit polynomial solution. The RMS of the residuals is $\sim70\,\mathrm{m\, s^{-1}}$ in the blue arm and $\sim55\,\mathrm{m\, s^{-1}}$ in the red arm. However, these fluctuations are random, and therefore will average out if a large number of H$_{2}$ transitions are used. Only systematic deviations from the true wavelength solution should appreciably affect the best estimate of $\Delta\mu/\mu$. There are fewer good ThAr lines in the blue end of the spectrum than in the red end, and therefore larger deviations of the wavelength solution from the true solution are possible. 

Following the analysis of \citet{Murphy:07b}, the systematic deviation in the blue end of the spectrum relative to the red end of the forest has an upper limit of $\sim20\,\mathrm{m\, s^{-1}}$. The maximum $K_{i}$ value used in the fit is 0.053, whilst the minimum is $-0.009$. This implies that the maximum possible systematic due to this effect is given by $\delta(\Delta\mu/\mu)=(\Delta v/c)/\Delta K_{i}$, which is $1.1\times10^{-6}$. In reality, the effect is likely to be smaller than this as positive deviations should tend to cancel somewhat with negative deviations. However, how to reduce the effect is unclear; it may not simply scale as $1/\sqrt{N}$ (for $N$ transitions). Therefore, we retain this estimate as the maximum possible systematic effect due to this cause.

\subsection{Intra-order distortions of unknown origin}

The path that the quasar light takes through the telescope is similar but not identical to that from the ThAr calibration lamp -- the ThAr light fills the slit nearly uniformly, whilst the quasar light does not, and the ThAr light does not pass through the telescope optics. Due to the different light paths, the wavelength scale of the quasar light may be different to that of the ThAr light; the differences between them may appear as an apparent distortion of the wavelength scale. Both long range and short-range distortions are possible.

\citet{Griest:09} identified a pattern of distortion within echelle orders in Keck/HIRES spectra, such that the wavelength scale at the centre of echelle orders is distorted with respect to that at the echelle order edges. The peak-to-peak velocity distortion is $\sim500\,\mathrm{m\, s^{-1}}$ at $\sim5600\mathrm{\AA}$. The distortion was identified by comparing the calibration of a spectrum obtained using a ThAr exposure to that obtained using an I$_{2}$ absorption cell. The iodine cell is placed in the quasar light path, and the characteristic absorption spectrum is imprinted on the quasar spectrum. The use of an iodine cell therefore obviates the concern about optical path differences when using a ThAr lamp. Unfortunately, an I$_{2}$ cell is not useful for calibration of general quasar observations, because the iodine transitions cover only a relatively narrow part of the optical range, and because of the loss of flux from the quasar as a result of the use of the cell. The observed distortion pattern appears to be dependent on wavelength, and the distortion may be larger at longer wavelengths. The precise origin of the distortions is unknown, and similarly it is unknown to what extent the distortions remain constant in time, and how they depend on extrinsic factors (e.g. telescope orientation, temperature, pressure and accuracy of quasar centering in the spectrograph slit). Therefore, it is not possible at present to adequately remove these distortions of the wavelength scale from observations.

\citet{Whitmore:10} identified a similar effect in VLT/UVES spectra, with a peak-to-peak velocity distortion of $\sim200\,\mathrm{m\, s^{-1}}$. The distortion appears to be much less consistent between echelle orders than that seen by \citeauthor{Griest:09}, however. Further observations by one of us (MTM) have shown that the observed wavelength distortion is definitely not constant over long periods of time i.e.\ more than several nights, which makes removal of the distortion extremely difficult.

Similar to \citet{Malec:10}, we attempted to estimate the magnitude of the error introduced into a determination of $\Delta\mu/\mu$ as a result of the observed velocity distortions. To do this, we used a triangular-shaped distortion, where the wavelengths of pixels at the centre of echelle orders were displaced by $+200\,\mathrm{m\, s^{-1}}$ with respect to those at the echelle order edges. The modification to the spectra was implemented within \textsc{uves\_popler}. The shift in $\Delta\mu/\mu$ after modifying the spectrum was $-0.3\times10^{-6}$ [using the 3-component, $b=F(J)$ model]. Clearly this value is model-dependent -- if the distortion has a different amplitude or form, then the impact on $\Delta\mu/\mu$ may be different. However, this estimate of the systematic error is likely to be of the correct magnitude. We therefore adopt a Gaussian with $\sigma=0.3\times10^{-6}$ as an estimate of the systematic effect due to distortions of this type.

\subsection{Velocity structure and spatial segregation}

It is possible that transitions arising from different $J$-levels might be spatially segregated \citep{Jenkins:97a,Levshakov:02a}. Assuming that all $J$-levels arise from the same redshift in this event could spuriously produce $\Delta\mu/\mu \neq 0$. Although our result for the Q0528$-$250 absorber is statistically consistent with zero, it is of course possible that a non-zero $\Delta\mu/\mu$ could be pushed towards zero by this sort of systematic effect. Similar to \citet{Malec:10}, we relaxed our assumption that corresponding components in all $J$-levels arise from the same redshift. In particular, we divided the data set into ``cold'' transitions, $J\in[0,1]$, and ``warm'' transitions, $J\in[2,4]$, as was done earlier, but only tie the redshifts of corresponding components between different $J$-levels within these two groups. If there is spatial segregation, this should be seen as a statistically significant difference between the redshifts of corresponding components between the two groups, and also as a substantial shift in the values of $\Delta\mu/\mu$ derived from the two groups compared to what was obtained earlier.

\begin{figure}
\includegraphics[bb=50 92 535 776,angle=-90,width=82.5mm]{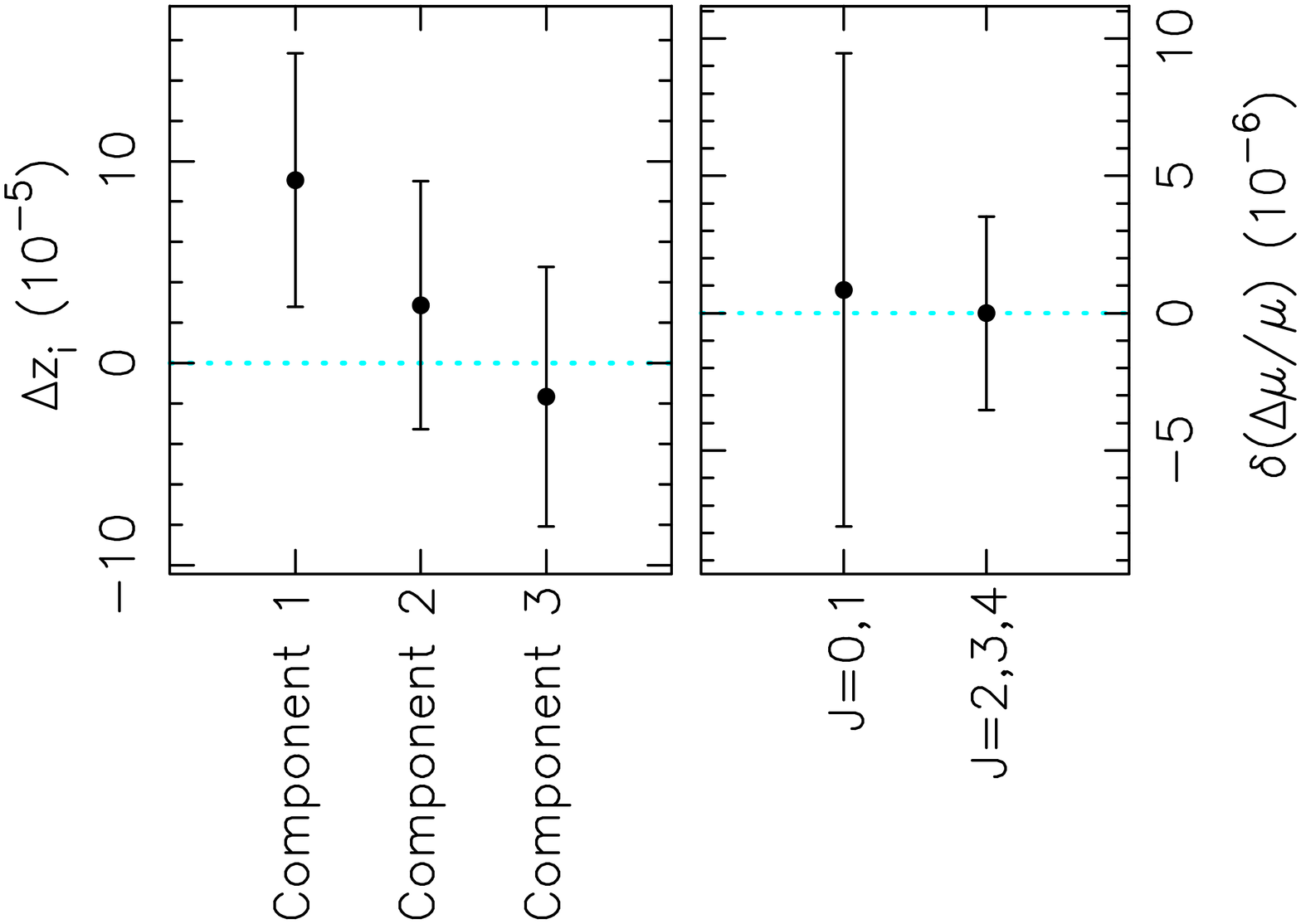}
  \caption{\emph{Left panel:} Difference between the redshifts of components for the H$_2$ fit for the 3-component model to the absorber when the velocity structure was allowed to vary between the ``cold'' and ``warm'' components. The difference for each component is defined as $z(J\in[0,1])-z(J\in[2,4])$. \emph{Right panel:} The difference in  $\Delta\mu/\mu$ when the velocity structure was allowed to vary, defined as $\Delta\mu/\mu(\mathrm{structure\ allowed\ to\ vary})-\Delta\mu/\mu(\mathrm{structure\ not\ allowed\ to\ vary})$. The error estimate is calculated as the mean of the error estimates in the two cases considered; the errors do not differ appreciably between the two cases. From consideration of both of these plots there does not appear to be any significant evidence that our assumptions about the velocity structure model have affected our estimate of $\Delta\mu/\mu$. \label{fig_Q0528_veldiff}}
\end{figure}

In the right panel of Fig.\  \ref{fig_Q0528_diff_J_contrib} we directly compared $\Delta\mu/\mu$ in the case where the velocity structure was allowed to vary between cold and warm components, and note that there is no appreciable shift. In Fig.\  \ref{fig_Q0528_veldiff} we show these considerations more directly by examining the differences in the redshifts of the three components, and also the explicit difference between $\Delta\mu/\mu$ in the two cases considered. We see that there is no statistically significant difference between the redshifts in any of the three components when the velocity structure is allowed to vary between the cold and warm components. Similarly, we see that the shift in $\Delta\mu/\mu$ is $\lesssim 0.1\sigma$ when the velocity structure is allowed to vary between the cold and warm components. Thus, we conclude that there is no evidence for a systematic shift in $\Delta\mu/\mu$ as a result of segregation of the cold and warm $J$-levels. 

Nevertheless, to quantify the possible error introduced by our assumptions regarding velocity structure, we examine the actual shifts in $\Delta\mu/\mu$. The shift in $\Delta\mu/\mu$ for the $J\in[0,1]$ levels from the original value [for the 3-component, $b=F(J)$ model] is $0.8\times10^{-6}$, and for the $J\in[2,4]$ levels is $0.2\times10^{-6}$. We thus take $0.8\times10^{-6}$ as an estimate of the potential error introduced into our analysis due to assumptions about the velocity structure in order to be conservative. 

\subsection{Re-dispersion of spectra}

The spectrum used is the result of the co-addition of exposures taken at different times, each with its own ThAr calibration. During the co-addition, the spectra are placed on a common wavelength grid. Because the spectra are re-binned, the choice of wavelength grid introduces correlations between neighbouring pixels. More importantly, the choice of the wavelength grid has the potential to affect $\Delta\mu/\mu$ by slightly distorting the absorption profile shapes. To investigate this, we examined the effect of shifting the wavelength grid by $-0.2$, $-0.1$, $0.1$ and $0.2$ pixels. The shifts this induced in $\Delta\mu/\mu$ from the original value are $-1.3\times10^{-6}$, $-2.1\times10^{-6}$, $+1.0\times10^{-6}$ and $+0.2\times10^{-6}$. The standard deviation of these values is $\approx1.4\times10^{-6}$, and so we adopt $1.4\times10^{-6}$ as an estimate of the potential error in $\Delta\mu/\mu$ on account of the re-dispersion of the contributing exposures.

\section{Result including systematic errors}\label{s_finalresult}

In Table \ref{tab_errorbudget}, we accumulate the potential systematic errors from our discussion above and give our final estimate of $\Delta\mu/\mu$ including the systematic component. Although the distribution of systematic errors is likely to be Gaussian in many cases, the impact on $\Delta\mu/\mu$ arising from the distortion in the wavelength scale (for instance) is an upper limit. The probability distribution of the sum of random variables is given by the convolution of their individual probability density functions. Thus, to estimate our final uncertainty, we convolve the distributions assumed for each of the sources of uncertainty, and give the standard deviation of the resultant distribution as our uncertainty estimate. 

\begin{table*}
 \begin{center}
\caption{Error budget for the analysis of the $z=2.811$ absorber toward Q0528$-$250 from the sources described in section \ref{s_systematics}. The third column gives the magnitude of the uncertainty estimate. The final error estimate is calculated as the standard deviation of the convolution of the assumed distributions of the individual error estimates; the assumed distributions are given in the fourth column.  \label{tab_errorbudget}}
\begin{tabular}{lccl}
\hline 
Source of error & $\Delta\mu/\mu$ ($10^{-6}$) & $\delta(\Delta\mu/\mu)$ ($10^{-6}$) & Assumed distribution \\ \hline
Systematic distortion of ThAr wavelength scale  &  & $\pm 1.1$ & Uniform\\
Intra-order wavelength scale distortions  &   & $\pm 0.3$ &  Gaussian\\
Velocity structure and spatial segregation  &   & $\pm 0.8$  & Gaussian\\
Re-dispersion of spectra &  & $\pm 1.4$ & Gaussian\\ \hline
Total systematic error &    & $\pm 1.9$ & \\ 
Statistical     &  0.3                           & $\pm 3.2$                             & Gaussian \\ \hline
Final estimate & 0.3 & $\pm 3.7$ & \\
\hline 
 \end{tabular}
 \end{center}
\end{table*}

This yields our final estimate of $\Delta\mu/\mu$ for the $z=2.811$ absorber toward Q0528$-$250 as
\begin{align}
\frac{\Delta\mu}{\mu} &= (0.3\pm3.2_{\mathrm{stat}}\pm1.9\mathrm{_{\mathrm{sys}})\times10^{-6}} \\
                      &= (0.3\pm3.7)_\mathrm{tot} \times10^{-6}. 
\end{align}

\section{Other analysis}

\subsection{H$_2$/HD column density ratio}

The HD/H$_2$ column density ratio is of astrophysical interest. We calculate the total H$_2$ column density under the 3-parameter, $b=F(J)$ model via the method given in the caption to Table \ref{tab_NJlevel} to be $\log_{10}(N/\mathrm{cm^{-2}})=16.556 \pm 0.024$. Column densities for each of the H$_2$ $J$-levels are given in Table \ref{tab_NJlevel}. The HD $J=0$ column density is $\log_{10}(N/\mathrm{cm^{-2}})=13.267 \pm 0.072$ directly from the fit. We assume that the total H$_2$/HD column density ratio is approximated by the total H$_2$/HD($J=0$) column density ratio \citep{Tumlinson:10a}. This directly gives the HD/H$_2$ column density ratio as $(5.4 \pm 1.1) \times 10^{-4}$, or $\log_{10}[N(\mathrm{HD})/N(\mathrm{H}_2)]= -3.3 \pm 0.2$. Alternatively, $\log_{10}[N(\mathrm{HD})/2N(\mathrm{H}_2)] = -3.6 \pm 0.2$. This compares with $\log_{10}[N(\mathrm{HD})/2N(\mathrm{H}_2)] = -4.1 \pm 0.2$ from the $z=2.059$ absorber toward J2123$-$0050 and $-4.8 \pm 1.5$ from the $z=2.627$ absorber toward FJ0812+32B \citep{Tumlinson:10a}. The result presented here for $\log_{10}[N(\mathrm{HD})/2N(\mathrm{H}_2)]$ from Q0528$-$250 differs from the value from J2123$-$0050 at the $1.8\sigma$ level, which is large but not grossly inconsistent, particularly since the estimate here is statistical only and does not consider gas cloud kinematics or potential systematic effects. Additionally, because the low-$J$ H$_2$ transitions in the Q0528$-$250 absorber are saturated it is difficult to accurately estimate the column density for H$_2$ in Q0528$-$250; \citet{Tumlinson:10a} noted this for the absorber in J2123$-$0050. As such, some caution is warranted in comparing the HD/H$_2$ ratio presented here to other results. Additionally, our fits were constructed with the H$_2$ line intensities fitted as free parameters. This makes it challenging to accurately estimate the total H$_2$ column density; to obtain a more accurate HD/H$_2$ column density ratio, this absorber should be re-analysed specifically for this purpose.

\begin{table}
 \begin{center}
\caption{Column densities for the different $J$-levels under the 3-component, $b=F(J)$ model. $n$ gives the number of transitions contributing to a particular $J$-level. Because the column densities for each transition are fitted as free parameters, the $\log_{10}$ column density for each $J$-level is calculated as the weighted mean of the $\log_{10}$ column densities for the transitions in that $J$-level, but where statistical errors are increased in quadrature with an additional term, $\sigma_J$, which is calculated so that $\chi^2_\nu=1$ about the weighted mean. Without the inclusion of $\sigma_J$, $\chi^2_\nu\gg1$, which reflects the accumulation of errors in the local model for the continuum and weak, unmodelled Lyman-$\alpha$ transitions which overlap with the transitions in question. $\sigma_J$ is given in the fourth column.  \label{tab_NJlevel}}
\begin{tabular}{lrcr}
\hline 
$J$     & $n$ & $\log_{10}(N/\mathrm{cm^{-2}})$  & $\log_{10}(\sigma_J)$ \\ \hline
0       & 3   & $15.68 \pm 0.25$                 & $0.43$ \\
1       & 8   & $16.028 \pm 0.094$               & $0.16$ \\
2       & 24  & $16.008 \pm 0.055$               & $0.23$ \\
3       & 28  & $15.917 \pm 0.030$               & $0.14$ \\
4       & 13  & $14.476 \pm 0.025$               & $0.084$ \\ \hline
Total   & 76  & $16.556 \pm 0.024$               & n/a\\
\hline 
 \end{tabular}
 \end{center}
\end{table}

\section{Discussion}

\begin{figure}
\includegraphics[bb=50 103 542 766,angle=-90,width=82.5mm]{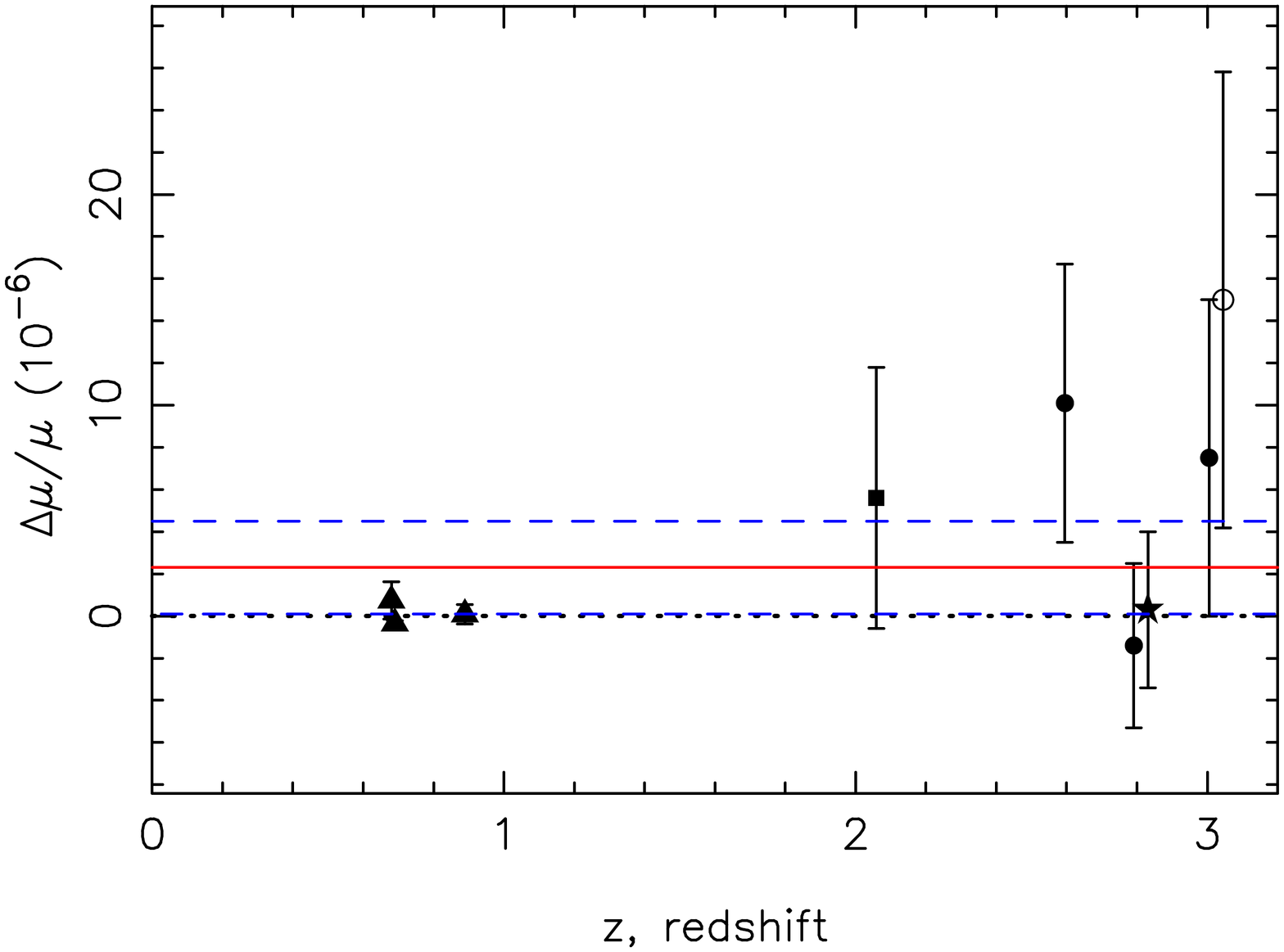}
  \caption{Recent cosmological constraints on $\Delta\mu/\mu$ derived from quasar absorption measurements. Filled circles indicate the H$_2$ measurements from Q0405$-$443, Q0347$-$383 and Q0528$-$250 in \citet{King:08}, the square indicates the H$_2$/HD measurements from J2123$-$0050 in \citet{Malec:10}, the star indicates the result of this paper, the open circle indicates the Q0347$-$383 measurement of \citet{Wendt:11a} [which is not independent of the Q0347$-$383 measurement in \citet{King:08} due to the use of common spectra] and the triangles indicate the constraints from ammonia in \citet{Murphy:Flambaum:08}, \citet{Henkel:09} and \citet{Kanekar:11a}. The two Q0528$-$250 points and the two Q0347$-$383 points have been slightly displaced in redshift for clarity. Where uncertainty estimates were given with both a statistical and systematic component, we added these in quadrature to derive the error bars shown. The red, solid line indicates the weighted mean of the $z>1$ points excluding the Q0347$-$383 point from \citet{Wendt:11a} [as it is not independent of the Q0347$-$383 point from \citet{King:08}], and the blue, dashed line indicates the $1\sigma$ uncertainty around that weighted mean. Clearly, if the $z<1$ measurements were included in the weighted mean then they would dominate the result due to their small error bars. \label{fig_all_H2_results}}
\end{figure}

Firstly, it is clear that the result obtained here is consistent with no variation of $\mu$. Secondly, the result is consistent with the Q0528$-$250 result in \citet{King:08}, and also the $\Delta\mu/\mu$ estimates obtained from Q0405$-$443 and Q0347$-$383 in \citet{King:08} and from J2123$-$0050 in \citet{Malec:10}. From consideration of all the H$_2$/HD measurements from quasar absorbers (which are all at $z>2$) there is no statistically significant evidence for cosmological variation in $\mu$. In particular, we can formally combine the results from \citet{King:08}, \citet{Malec:10} and this work under a weighted mean model (assuming that a weighted mean model is a legitimate description of the data), and we obtain $\Delta\mu/\mu = (2.3\pm 2.2)\times 10^{-6}$, with $\chi^2_\nu$ about the weighted mean of $\approx 0.9$. The fact that $\chi^2_\nu \approx 1$ suggests that the uncertainty estimates on $\Delta\mu/\mu$ are approximately correct. We show the result of this work in comparison with other quasar constraints on $\Delta\mu/\mu$ in Fig.\   \ref{fig_all_H2_results}.

\section{Conclusion}

We have used new observations of the molecular absorber at $z=2.811$ towards Q0528$-$250 to derive a new, strong constraint on $\Delta\mu/\mu$ at high redshift, namely $\Delta\mu/\mu = (0.3\pm3.2_{\mathrm{stat}}\pm1.9\mathrm{_{\mathrm{sys}})\times10^{-6}}=(0.3\pm3.7)\times10^{-6}$. This result is the strongest individual constraint on variation in $\mu$ at $z>1$, and together with other measurements of H$_2$/HD absorbers provides a very strong constraint on evolution in $\mu$ over most of the observable history of the universe. The value of $\Delta\mu/\mu$ presented here is consistent with the values presented in \citet{King:08} and \citet{Malec:10}.

Our result was derived using 76 H$_2$ transitions and seven HD transitions. This is the first time that HD has been detected in this absorber, with a column density of $\log_{10}(N/\mathrm{cm^{-2}})=13.27 \pm 0.07$. We measure the H$_2$/HD column density ratio to be $\log_{10}[N(\mathrm{HD})/N(\mathrm{H}_2)] = -3.3 \pm 0.2$.

Similar to \citet{King:08} and \citet{Malec:10}, we applied the comprehensive fitting method (CFM), where we simultaneously fitted the H$_2$/HD transitions with the surrounding Lyman-$\alpha$ forest lines. By modelling the forest simultaneously with the H$_2$/HD transitions, our estimate of $\Delta\mu/\mu$ should be more accurate. Similarly, the uncertainty in determining the locations of the forest lines directly feeds into our uncertainty estimate on $\Delta\mu/\mu$, helping to ensure that our uncertainty is not under-estimated. 

On the basis of the AICC, we concluded that a model of the absorber with 3 velocity components is preferred. We imposed a number of assumptions about the velocity structure of the absorber in order to reduce the number of free parameters in the fit. Investigation of potential differences between `cold' ($J \in [0,1]$) and `warm' ($J \in [2,4]$) components showed that the potential impact of our assumptions on the estimate of $\Delta\mu/\mu$ is significantly smaller than the statistical error on $\Delta\mu/\mu$. 
 
We investigated a number of potential systematic effects, including: known wavelength calibration errors due to uncertainties in the ThAr calibration; observed intra-echelle order wavelength scale distortions of unknown origin; assumptions about the velocity structure, and; the effect of re-dispersing the individual exposures onto a common wavelength grid. The contribution of each of these effects is small, and in aggregate the total uncertainty in $\Delta\mu/\mu$ is dominated by statistical rather than systematic error sources. This is reassuring, and means that future VLT/UVES observations of H$_2$/HD absorbers may continue to increase the constraint on $\Delta\mu/\mu$ in a meaninful way, provided that appropriate absorbers can be identified. 

Although there is no significant evidence for evolution in $\mu$, and little ($\sim 2\sigma$) evidence for dipolar variation in $\mu$ across the sky, it is intriguing that the dipole direction in a dipole+monopole model fitted to the $z>1$ H$_2$/HD $\Delta\mu/\mu$ constraints points in a similar direction to the $\Delta\alpha/\alpha$ dipole described in \citet{King:11a}. Given the distribution of existing H$_2$ results on the sky, new H$_2$ absorbers in judicious locations will be able to place strong constraints on spatial variation in $\mu$. 

The current number of quasars known to contain molecular hydrogen absorption is small, and almost all existing analyses have focused on the objects Q0405$-$443, Q0347$-$383, Q0528$-$250 and J2123$-$0050. Given that analysis of molecular hydrogen currently remains the most direct and precise way of investigating potential evolution in $\mu$ at $z>2$, it would be advantageous if many more H$_2$/HD systems could be discovered. Although new observations of existing H$_2$/HD systems can increase the SNR, common systematics may remain. The use of more absorbers in different quasars would increase both the redshift and spatial coverage of the $\Delta\mu/\mu$ constraints.

\section*{Acknowledgments}

This work is based on observations carried out at the European Southern Observatory (ESO) under program ID 82.A-0087 (PI Ubachs), with the UVES spectrograph installed at the Kueyen UT2 on Cerro Paranal, Chile. The authors thank L. Kaper (Amsterdam), V. Flambaum (UNSW) and J. Berengut (UNSW) for fruitful discussions. JAK was supported in part by an Australian Postgraduate Award. JAK additionally thanks VU for support to travel to Amsterdam, where part of this work was carried out. MTM thanks the Australian Research Council for a QEII Research Fellowship (DP0877998). WU acknowledges support from the Netherlands Foundation for Fundamental Research of Matter (FOM).

\bsp

\bibliographystyle{mn2e.bst}
\bibliography{Q0528_B2.bib}

\appendix

\section{Voigt profile fits}\label{app_voigt_profile_fits}

\begin{figure*}
\includegraphics[bb=62 181 544 801,width=160mm]{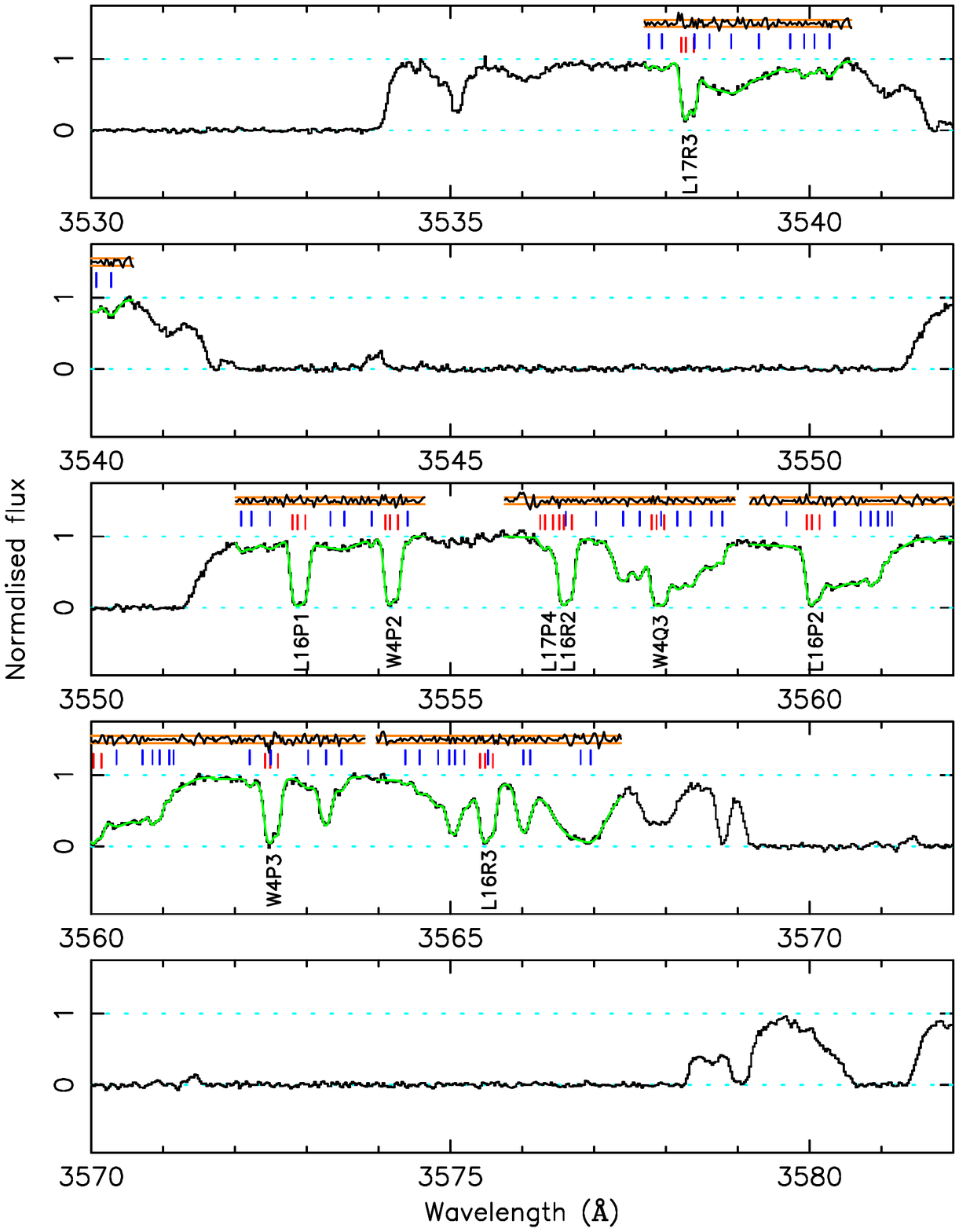}
  \caption{H$_2$/HD fit for the $z=2.811$ absorber toward Q0528$-$250 (part 1). The vertical axis shows normalised flux. The model fitted to the spectrum is shown in green. Red tick marks indicate the position of H$_2$/HD components, whilst blue tick marks indicate the position of blending transitions (presumed to be Lyman-$\alpha$). Normalised residuals (i.e. [data - model]/error) are plotted above the spectrum between the orange bands, which represent $\pm 1\sigma$. Labels for the H$_2$ transitions are plotted below the data.\label{fig_specplot_start}}
\end{figure*}

\begin{figure*}
\includegraphics[bb=62 181 544 801,width=160mm]{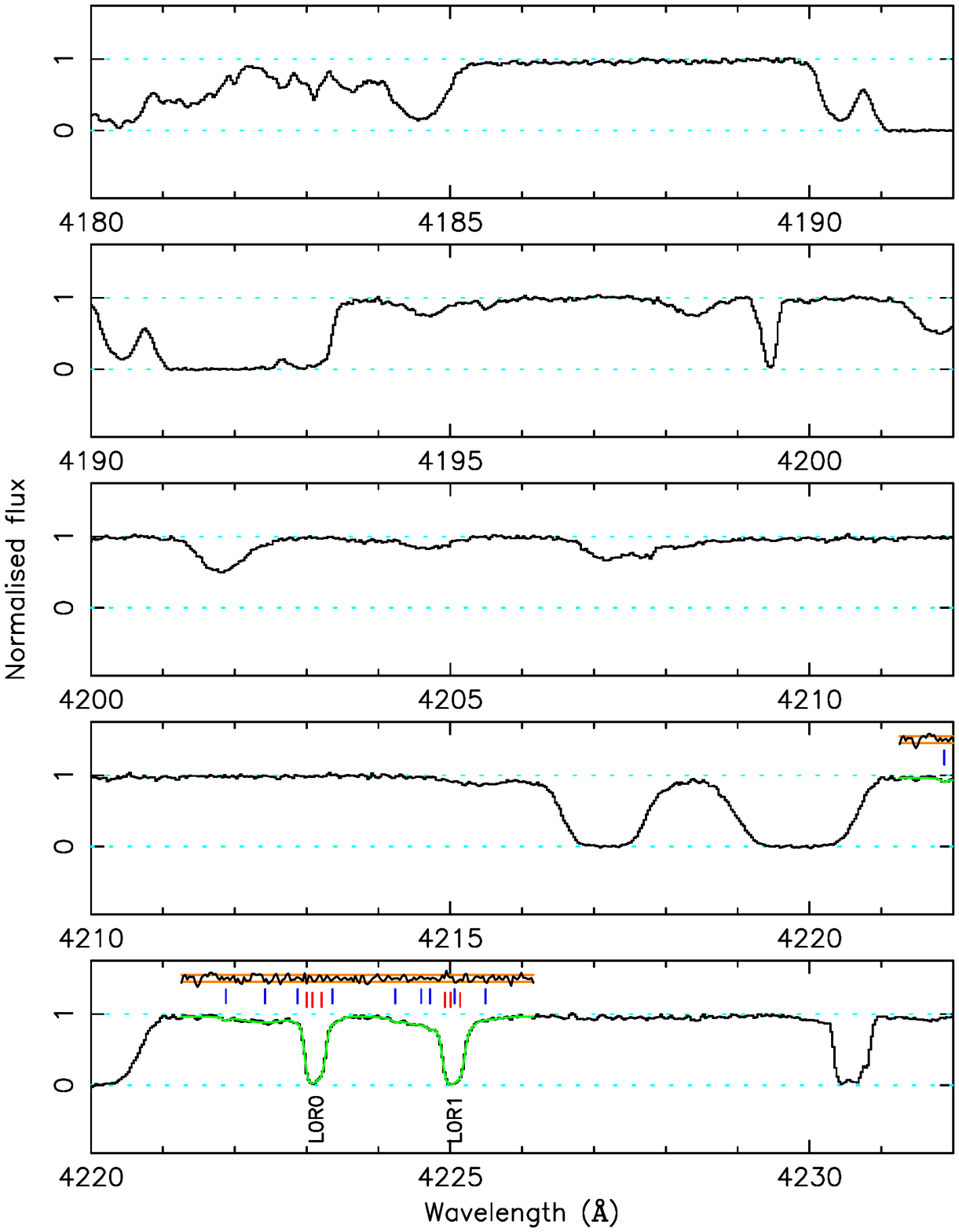}
  \caption{H$_2$/HD fit for the $z=2.811$ absorber toward Q0528$-$250 (part 14). The vertical axis shows normalised flux. The model fitted to the spectrum is shown in green. Red tick marks indicate the position of H$_2$/HD components, whilst blue tick marks indicate the position of blending transitions (presumed to be Lyman-$\alpha$). Normalised residuals (i.e. [data - model]/error) are plotted above the spectrum between the orange bands, which represent $\pm 1\sigma$. Labels for the H$_2$ transitions are plotted below the data.\label{fig_specplot_end}}
\end{figure*}

\emph{Online only}: Figures \ref{fig_specplot_start} through \ref{fig_specplot_end} show our Voigt profile model for the $z=2.811$ absorber toward Q0528$-$250 and the surrounding Lyman-$\alpha$ forest regions, indicating both the positions of the H$_2$/HD components as well as the H\,\iscs components used to fit the surrounding Lyman-$\alpha$ forest.

\begin{figure*}
\includegraphics[bb=62 181 544 801,width=160mm]{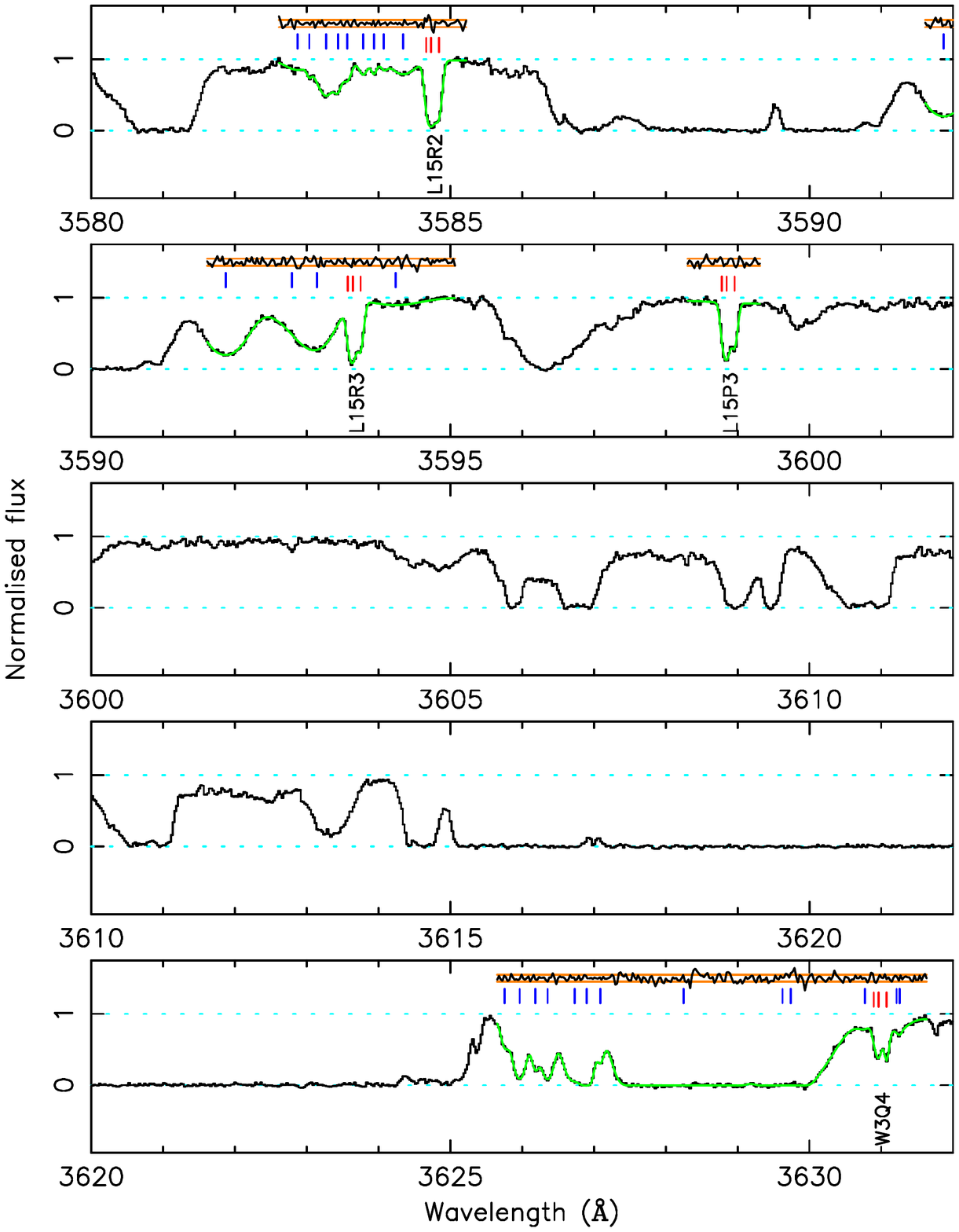}
  \caption{H$_2$/HD fit for the $z=2.811$ absorber toward Q0528$-$250 (part 2). The vertical axis shows normalised flux. The model fitted to the spectrum is shown in green. Red tick marks indicate the position of H$_2$/HD components, whilst blue tick marks indicate the position of blending transitions (presumed to be Lyman-$\alpha$). Normalised residuals (i.e. [data - model]/error) are plotted above the spectrum between the orange bands, which represent $\pm 1\sigma$. Labels for the H$_2$ transitions are plotted below the data.}
\end{figure*}
\begin{figure*}
\includegraphics[bb=62 181 544 801,width=160mm]{images/specplots/H2plot_p3.pdf}
  \caption{H$_2$/HD fit for the $z=2.811$ absorber toward Q0528$-$250 (part 3). The vertical axis shows normalised flux. The model fitted to the spectrum is shown in green. Red tick marks indicate the position of H$_2$/HD components, whilst blue tick marks indicate the position of blending transitions (presumed to be Lyman-$\alpha$). Normalised residuals (i.e. [data - model]/error) are plotted above the spectrum between the orange bands, which represent $\pm 1\sigma$. Labels for the H$_2$ transitions are plotted below the data.}
\end{figure*}
\begin{figure*}
\includegraphics[bb=62 181 544 801,width=160mm]{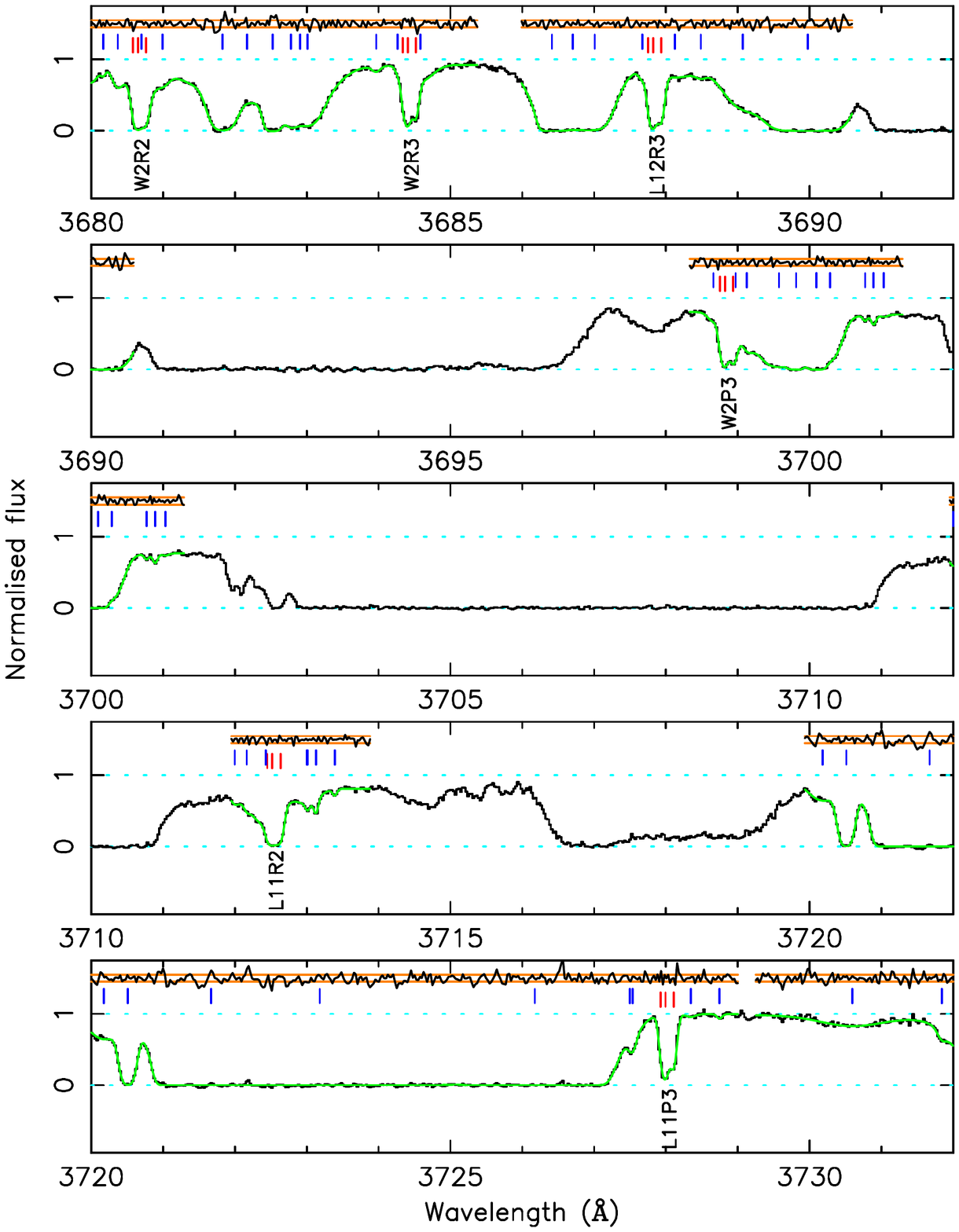}
  \caption{H$_2$/HD fit for the $z=2.811$ absorber toward Q0528$-$250 (part 4). The vertical axis shows normalised flux. The model fitted to the spectrum is shown in green. Red tick marks indicate the position of H$_2$/HD components, whilst blue tick marks indicate the position of blending transitions (presumed to be Lyman-$\alpha$). Normalised residuals (i.e. [data - model]/error) are plotted above the spectrum between the orange bands, which represent $\pm 1\sigma$. Labels for the H$_2$ transitions are plotted below the data.}
\end{figure*}
\begin{figure*}
\includegraphics[bb=62 181 544 801,width=160mm]{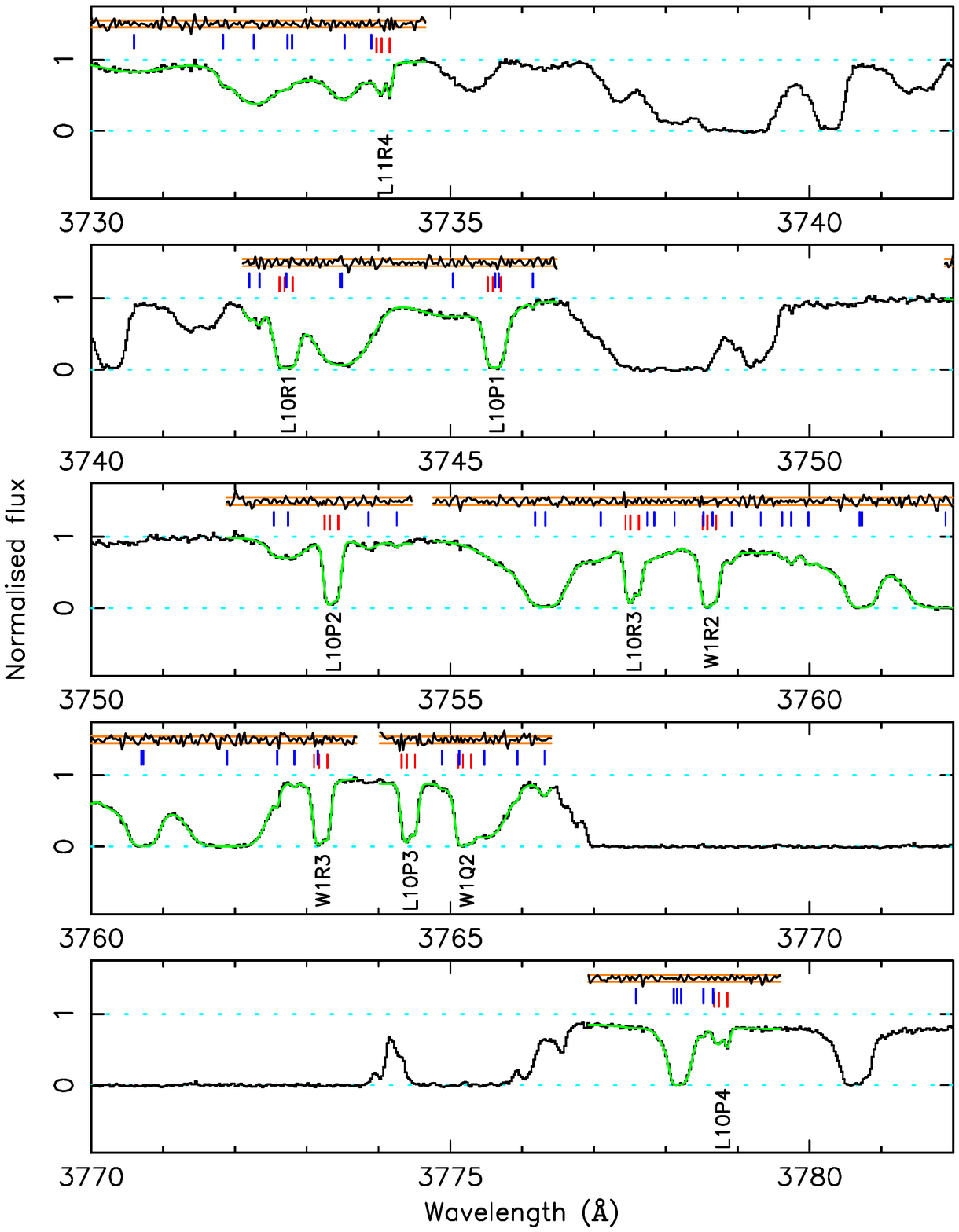}
  \caption{H$_2$/HD fit for the $z=2.811$ absorber toward Q0528$-$250 (part 5). The vertical axis shows normalised flux. The model fitted to the spectrum is shown in green. Red tick marks indicate the position of H$_2$/HD components, whilst blue tick marks indicate the position of blending transitions (presumed to be Lyman-$\alpha$). Normalised residuals (i.e. [data - model]/error) are plotted above the spectrum between the orange bands, which represent $\pm 1\sigma$. Labels for the H$_2$ transitions are plotted below the data.}
\end{figure*}
\begin{figure*}
\includegraphics[bb=62 181 544 801,width=160mm]{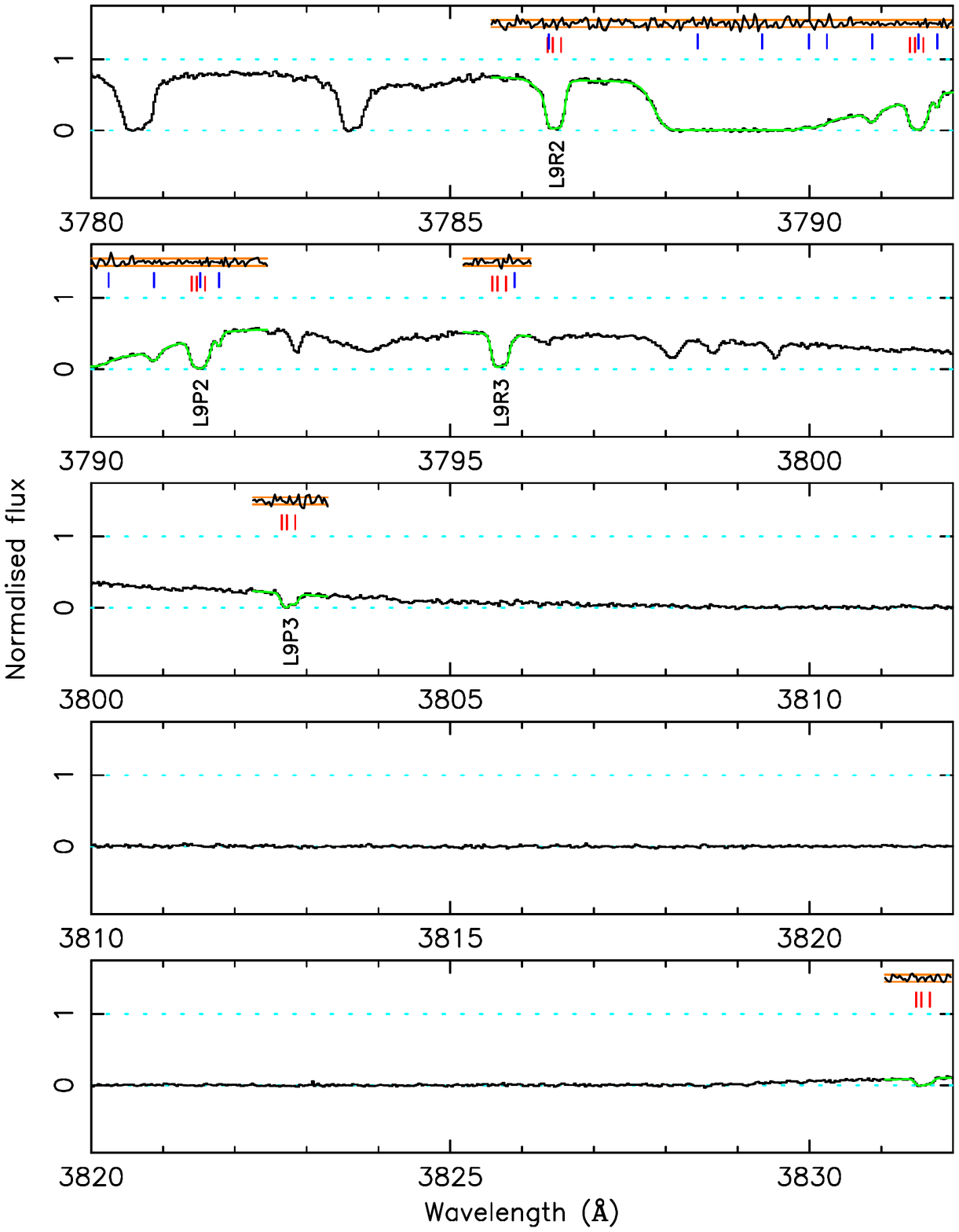}
  \caption{H$_2$/HD fit for the $z=2.811$ absorber toward Q0528$-$250 (part 6). The vertical axis shows normalised flux. The model fitted to the spectrum is shown in green. Red tick marks indicate the position of H$_2$/HD components, whilst blue tick marks indicate the position of blending transitions (presumed to be Lyman-$\alpha$). Normalised residuals (i.e. [data - model]/error) are plotted above the spectrum between the orange bands, which represent $\pm 1\sigma$. Labels for the H$_2$ transitions are plotted below the data.}
\end{figure*}
\begin{figure*}
\includegraphics[bb=62 181 544 801,width=160mm]{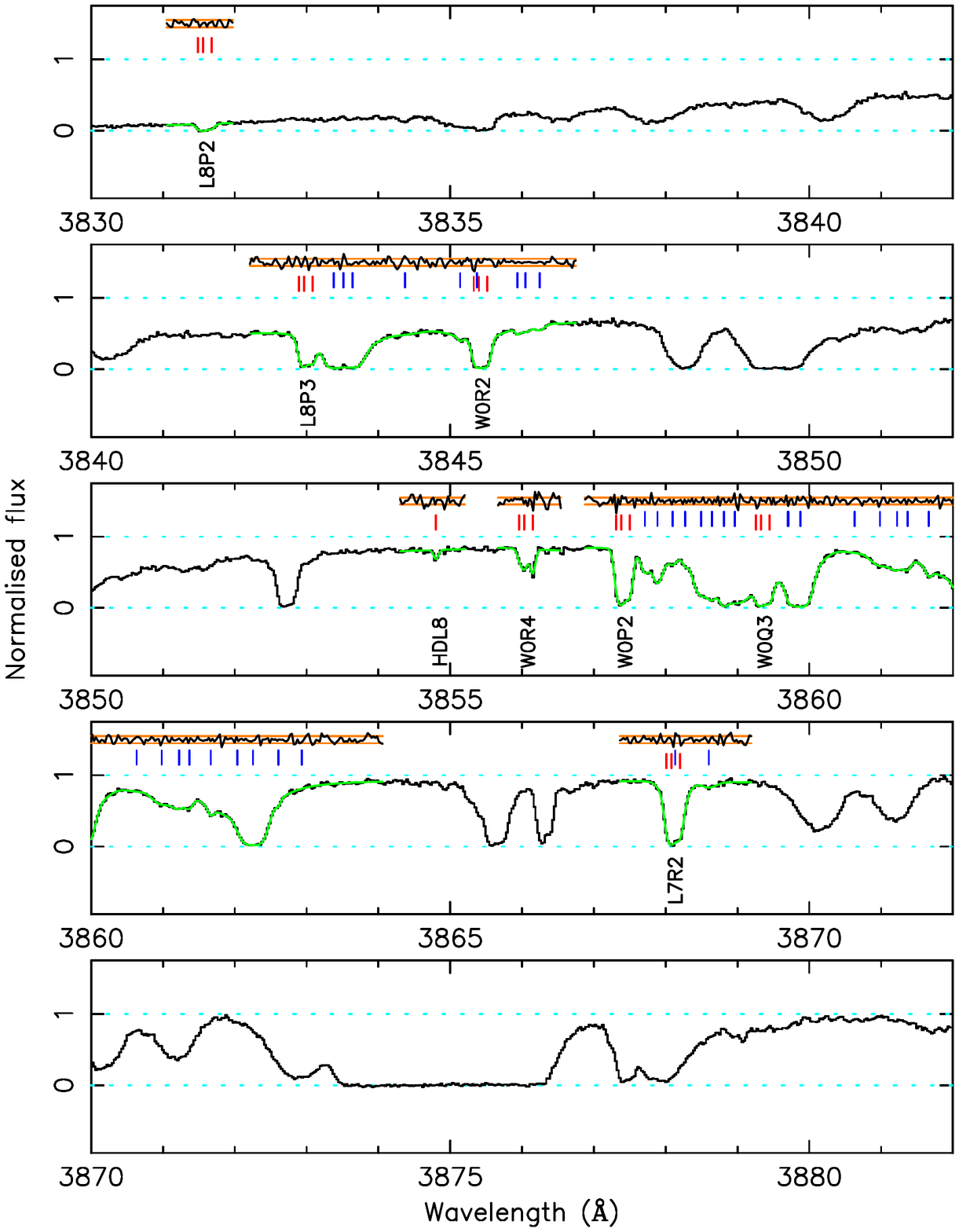}
  \caption{H$_2$/HD fit for the $z=2.811$ absorber toward Q0528$-$250 (part 7). The vertical axis shows normalised flux. The model fitted to the spectrum is shown in green. Red tick marks indicate the position of H$_2$/HD components, whilst blue tick marks indicate the position of blending transitions (presumed to be Lyman-$\alpha$). Normalised residuals (i.e. [data - model]/error) are plotted above the spectrum between the orange bands, which represent $\pm 1\sigma$. Labels for the H$_2$ transitions are plotted below the data.}
\end{figure*}
\begin{figure*}
\includegraphics[bb=62 181 544 801,width=160mm]{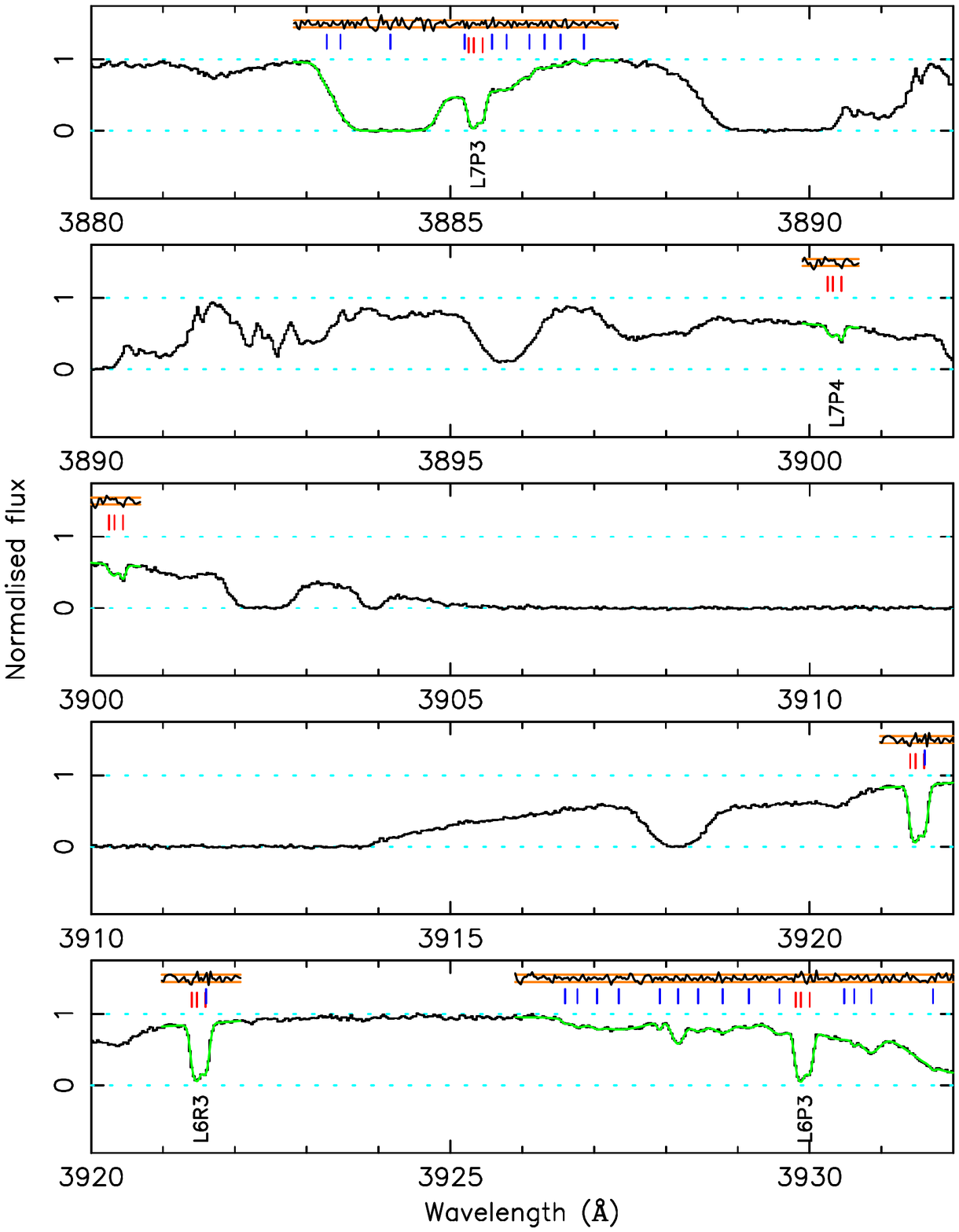}
  \caption{H$_2$/HD fit for the $z=2.811$ absorber toward Q0528$-$250 (part 8). The vertical axis shows normalised flux. The model fitted to the spectrum is shown in green. Red tick marks indicate the position of H$_2$/HD components, whilst blue tick marks indicate the position of blending transitions (presumed to be Lyman-$\alpha$). Normalised residuals (i.e. [data - model]/error) are plotted above the spectrum between the orange bands, which represent $\pm 1\sigma$. Labels for the H$_2$ transitions are plotted below the data.}
\end{figure*}
\begin{figure*}
\includegraphics[bb=62 181 544 801,width=160mm]{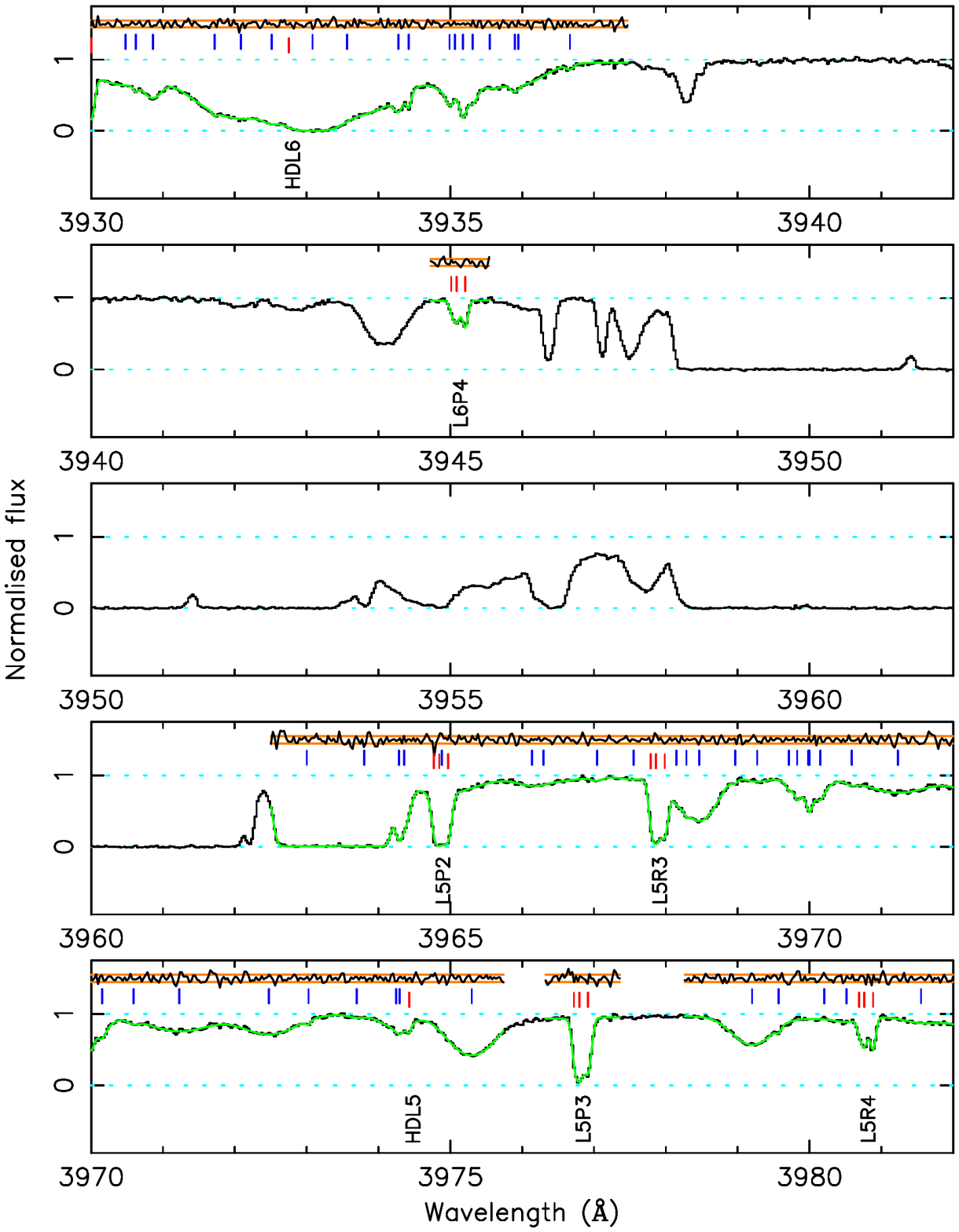}
  \caption{H$_2$/HD fit for the $z=2.811$ absorber toward Q0528$-$250 (part 9). The vertical axis shows normalised flux. The model fitted to the spectrum is shown in green. Red tick marks indicate the position of H$_2$/HD components, whilst blue tick marks indicate the position of blending transitions (presumed to be Lyman-$\alpha$). Normalised residuals (i.e. [data - model]/error) are plotted above the spectrum between the orange bands, which represent $\pm 1\sigma$. Labels for the H$_2$ transitions are plotted below the data.}
\end{figure*}
\begin{figure*}
\includegraphics[bb=62 181 544 801,width=160mm]{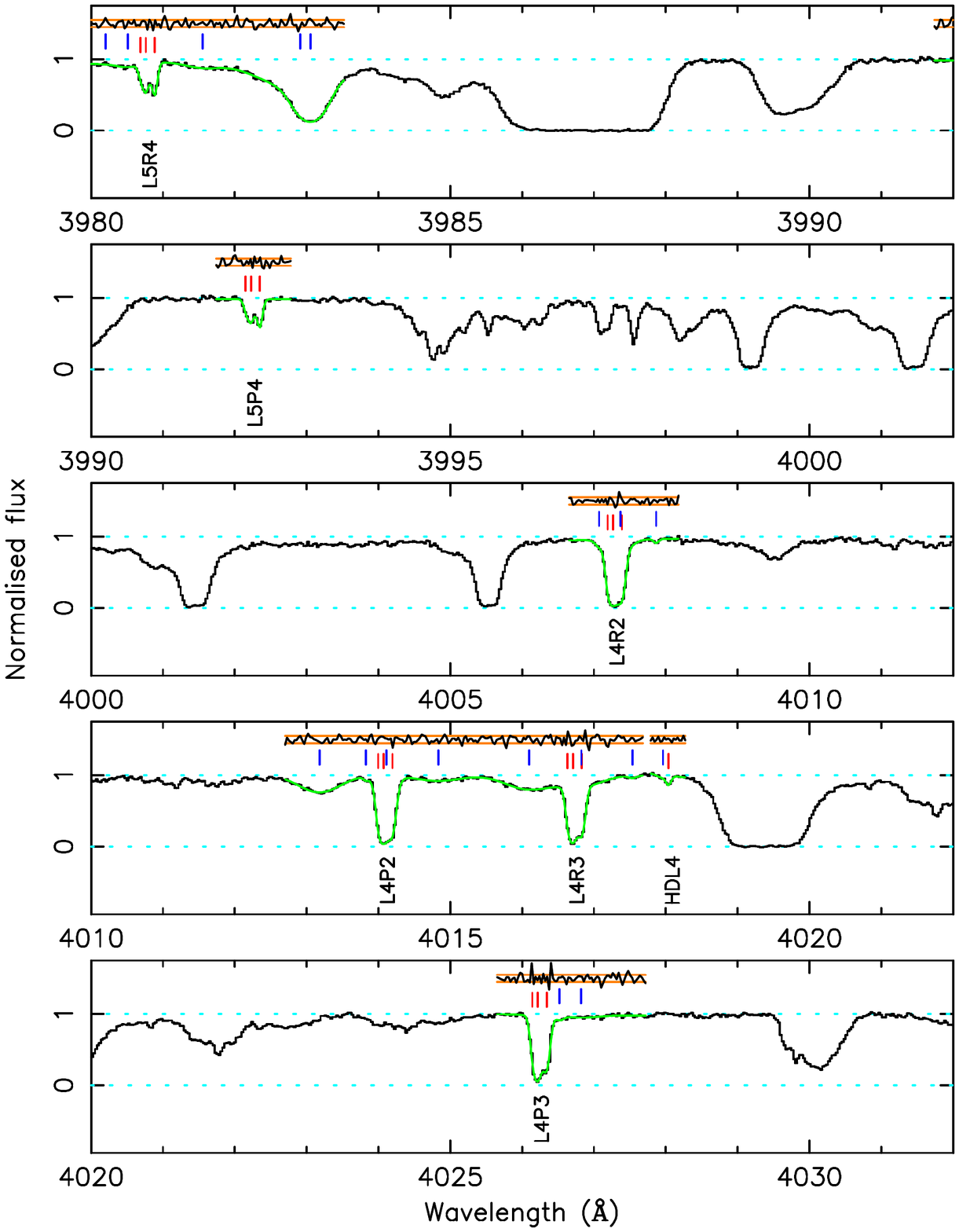}
  \caption{H$_2$/HD fit for the $z=2.811$ absorber toward Q0528$-$250 (part 10). The vertical axis shows normalised flux. The model fitted to the spectrum is shown in green. Red tick marks indicate the position of H$_2$/HD components, whilst blue tick marks indicate the position of blending transitions (presumed to be Lyman-$\alpha$). Normalised residuals (i.e. [data - model]/error) are plotted above the spectrum between the orange bands, which represent $\pm 1\sigma$. Labels for the H$_2$ transitions are plotted below the data.}
\end{figure*}
\begin{figure*}
\includegraphics[bb=62 181 544 801,width=160mm]{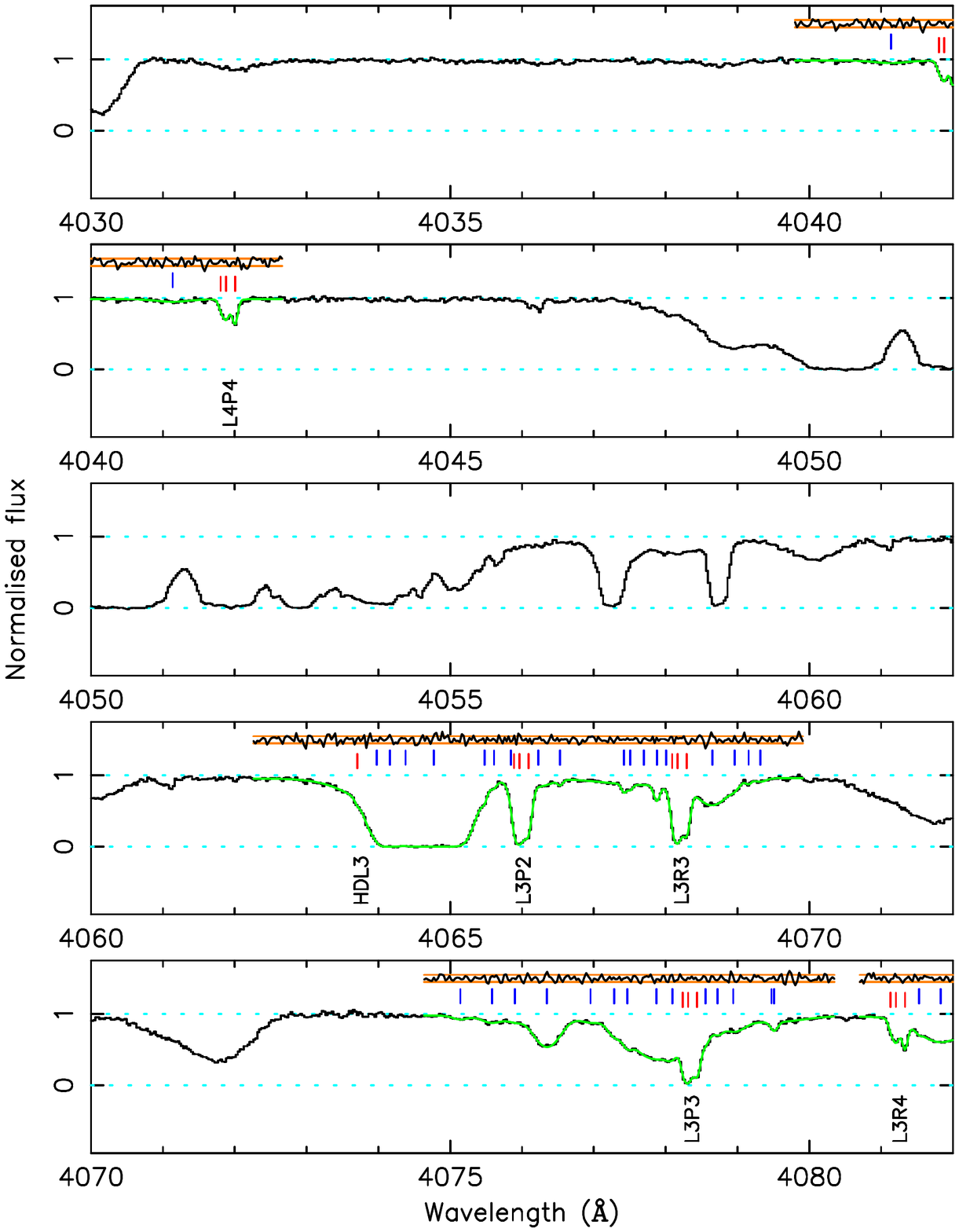}
  \caption{H$_2$/HD fit for the $z=2.811$ absorber toward Q0528$-$250 (part 11). The vertical axis shows normalised flux. The model fitted to the spectrum is shown in green. Red tick marks indicate the position of H$_2$/HD components, whilst blue tick marks indicate the position of blending transitions (presumed to be Lyman-$\alpha$). Normalised residuals (i.e. [data - model]/error) are plotted above the spectrum between the orange bands, which represent $\pm 1\sigma$. Labels for the H$_2$ transitions are plotted below the data.}
\end{figure*}
\begin{figure*}
\includegraphics[bb=62 181 544 801,width=160mm]{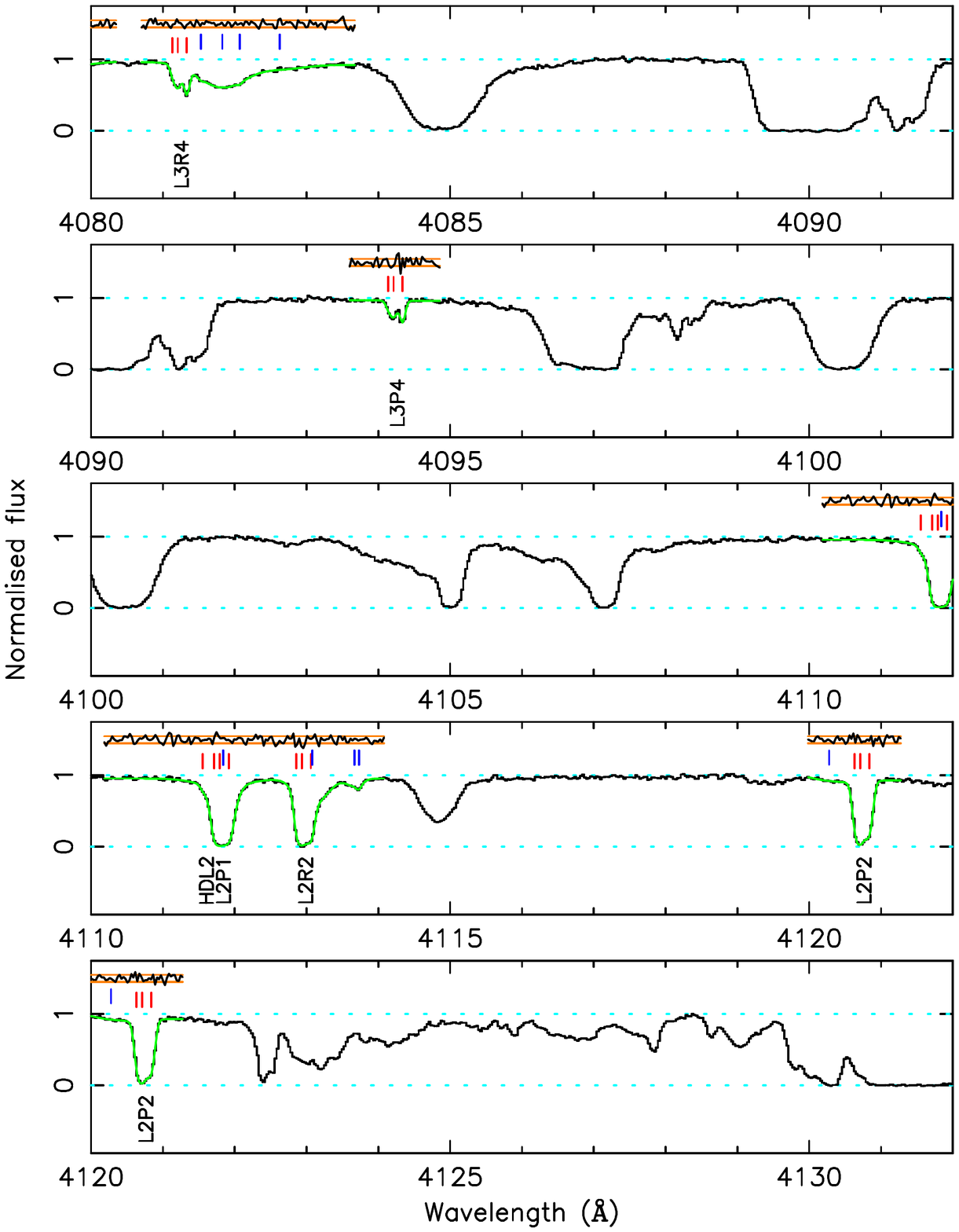}
  \caption{H$_2$/HD fit for the $z=2.811$ absorber toward Q0528$-$250 (part 12). The vertical axis shows normalised flux. The model fitted to the spectrum is shown in green. Red tick marks indicate the position of H$_2$/HD components, whilst blue tick marks indicate the position of blending transitions (presumed to be Lyman-$\alpha$). Normalised residuals (i.e. [data - model]/error) are plotted above the spectrum between the orange bands, which represent $\pm 1\sigma$. Labels for the H$_2$ transitions are plotted below the data.}
\end{figure*}
\begin{figure*}
\includegraphics[bb=62 181 544 801,width=160mm]{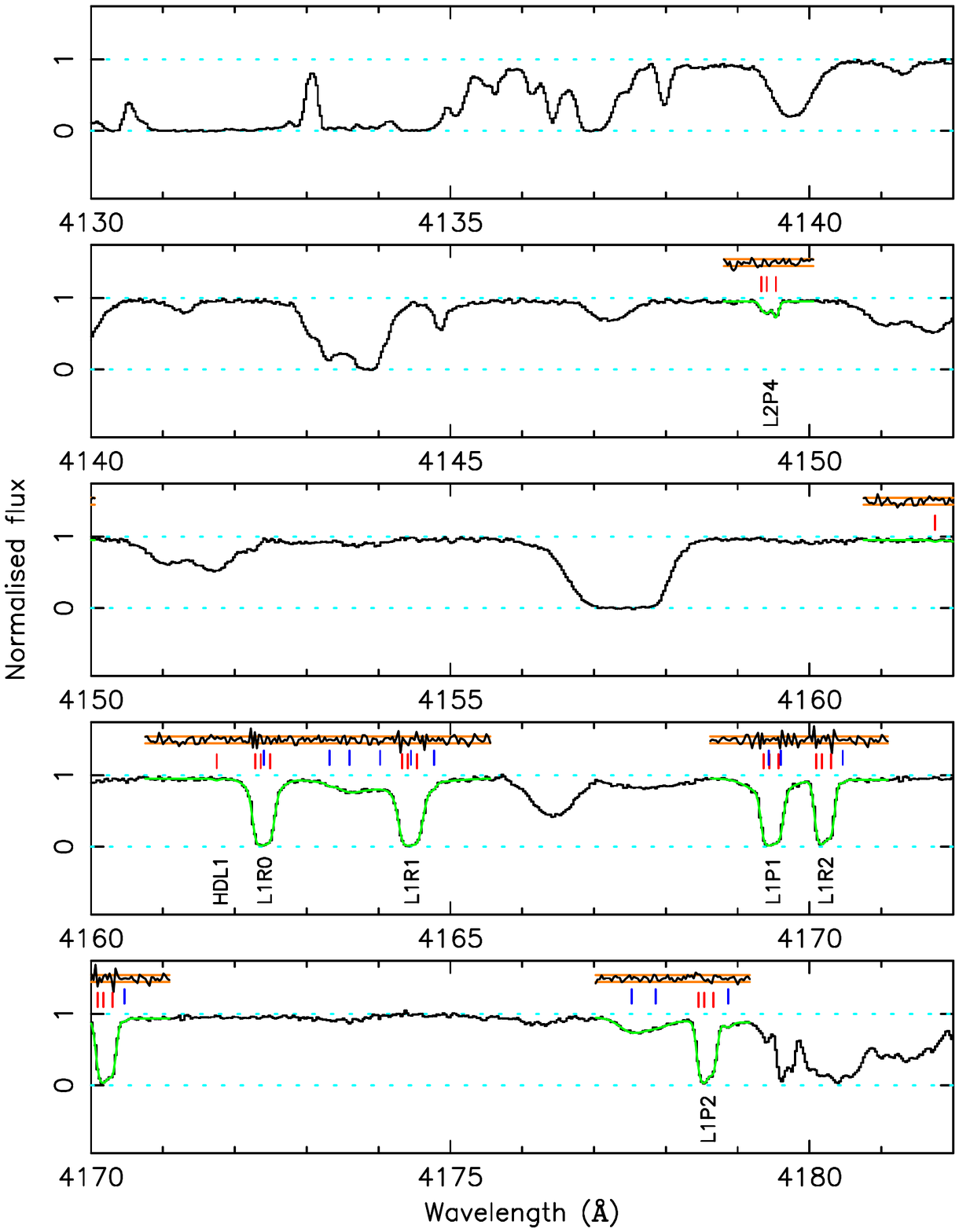}
  \caption{H$_2$/HD fit for the $z=2.811$ absorber toward Q0528$-$250 (part 13). The vertical axis shows normalised flux. The model fitted to the spectrum is shown in green. Red tick marks indicate the position of H$_2$/HD components, whilst blue tick marks indicate the position of blending transitions (presumed to be Lyman-$\alpha$). Normalised residuals (i.e. [data - model]/error) are plotted above the spectrum between the orange bands, which represent $\pm 1\sigma$. Labels for the H$_2$ transitions are plotted below the data.}
\end{figure*}

\label{lastpage}

\end{document}